# Cloud Storage Forensics: Analysis of Data Remnants on SpiderOak, JustCloud, and pCloud

SeyedHossein Mohtasebi[1], Ali Dehghantanha[2*], Kim Kwang Raymond Choo[3]

1- *Shabakeh Gostar Ltd. Co., Tehran, Iran,* shmohtasebi@gmail.com
2- School of Computing, Science and Engineering, University of Salford, Greater Manchester, M5 4WT, United Kingdom, A.Dehghantanha@salford.ac.uk
   *Corresponding Author: 215- Newton Building, School of Computing, Science and Engineering, University of Salford, Greater Manchester, M5 4WT, United Kingdom, A.Dehghantanha@salford.ac.uk, Tel: +44 (0)161 295 3531. A.Dehghantanha@Salford.ac.uk
3- Kim-Kwang Raymond Choo, Information Assurance Research Group, Advanced Computing Research Center, University of South-Australia, Mawson Lake Campus, Mawson Lake Boulevard, Mason Lakes, SA 5095, Australia, Raymond.Choo@unisa.edu.au.

**Abstract**: STorage as a Service (STaaS) cloud platforms benefits such as getting access to data anywhere, anytime, on a wide range of devices made them very popular among businesses and individuals. As such forensics investigators are increasingly facing cases that involve investigation of STaaS platforms. Therefore, it is essential for cyber investigators to know how to collect, preserve, and analyse evidences of these platforms. In this paper, we describe investigation of three STaaS platforms namely SpiderOak, JustCloud, and pCloud on Windows 8.1 and iOS 8.1.1 devices. Moreover, possible changes on uploaded and downloaded files metadata on these platforms would be tracked and their forensics value would be investigated.

Keywords: Digital Forensics, Cloud Computing Forensics, Cloud Storage Forensics, SpiderOak Forensics, JustCloud Forensics, pCloud Forensics

# 1. Introduction

Cloud computing enables businesses and individual to access computing resources such as servers and storages on an on-demand basis, where resources can be quickly provisioned with minimal efforts and interaction with the service provider (Mell & Grance, 2011). The National Institute of Standards and Technology (NIST) broadly categorised cloud computing services into three categories:

1. Software as a Service (SaaS): When an application is used to access shared infrastructure of the Cloud Storage Service Provider (CSSP). A popular example of SaaS is Storage-as-a-Service (STaaS) cloud systems.
2. Platform as a Service (PaaS): Users may deploy their own applications on the CSSP's infrastructure
3. Infrastructure as a Service (IaaS): The CSSP provides the underlying computing resources for the deployment of software (including Operation System (OS)) by the users (Mell & Grance, 2011).

The increasing popularity of STaaS and the potential for such services to be criminally exploited have attracted the attention of policing and forensics scholars in recent years (Damshenas, Dehghantanha, Mahmoud, & bin Shamsuddin, 2012). Many emerging and new platforms are now based on cloud services to operate. As noted by the National Institute of Standards and Technology (2014), cloud forensics is the application of scientific rules, technological exercises and approved methods to rebuild past events of crime committed in the cloud computing environment.

A number of researchers have reported data collection and preservation (A Aminnezhad, Dehghantanha, Abdullah, & Damshenas, 2013; Damshenas, Dehghantanha, Mahmoud, & Shamsuddin, 2014; F Daryabar & Dehghantanha, 2014; Hogben & Dekker, 2012; Hooper, Martini, & Choo, 2013; D Quick, Martini, & Choo, 2014), cloud malware detection (Damshenas, Dehghantanha, & Mahmoud, 2013; Farid Daryabar & Ali Dehghantanha, 2011) analysis and even privacy (Asou Aminnezhad & Ali Dehghantanha, 2012; Farid Daryabar, Dehghantanha, Udzir, & Fazlida, 2013; Dehghantanha & Franke, 2014) as main challenges during investigation of cloud platforms. As explained by Martini and Choo (2014a), "the large number of open challenges presented in a report by the National Institute of Standards and Technology (2014) demonstrates that, at the time of the research, cloud forensics remains an unresolved and somewhat under researched area of enquiry".

Although a number of STaaS have been examined (see Table 1), researchers (K.-K. R. Choo & Smith, 2008; K.-K. R. Choo, 2008) have posited that organized crime and cyber criminals are innovative and will constantly seek to "innovate" in order to evade the scrutiny and reach of law enforcement, such as using other STaaS to store incriminating evidence.

Table 1: A snapshot of existing cloud forensics research (Adapted from Martini and Choo (2012))

| *Model* | *Public cloud* | *Private cloud* |
|---|---|---|
| SaaS (including STaaS) | Amazon S3, Dropbox, Evernote and Google Docs (Chung, Park, Lee, & Kang, 2012) Dropbox, Google Drive and Microsoft SkyDrive (Federici, 2014) Dropbox, Box, and SugarSync (Grispos, Glisson, & Storer, 2013) Amazon Cloud Drive (Hale, 2013) Dropbox, OneDrive, and ownCloud (Martini, Do, & Choo, n.d.) Google Document, Flicker, PicasaWeb (Marturana, Me, & Tacconi, 2012) iCloud (Oestreicher, 2014) Google Drive (Darren Quick & Choo, 2014) Dropbox (Darren Quick & Choo, 2013b) SkyDrive (Darren Quick & Choo, 2013a) Google Drive, Dropbox, and SkyDrive (Darren Quick & Choo, 2013c) UbuntuOne (M Shariati, Dehghantanha, Martini, & Choo, n.d.; Mohammad Shariati, Dehghantanha, Martini, & Choo, 2015) hubiC (Blakeley, Cooney, Dehghantanha, & Aspin, 2015) SugarSync (Mohammad Shariati, Dehghantanha, & Choo, 2016) Mega (F. Daryabar, A. Dehghantanha & R. Choo, 2016) OneDrive, Box, GoogleDrive and Dropbox (F.Daryabar; A. Dehghantanha; B. Eterovic-Soric & R. Choo, 2016) | Dropbox, Box, and SugarSync (Grispos et al., 2013) ownCloud (Martini & Choo, 2013) |
| IaaS | | Hadoop (Cho, Chin, & Chung, 2012) Amazon EC2 (Dykstra & Sherman, 2012) vCloud (Martini & Choo, 2014c) XtreemFS (Martini & Choo, 2014b) Eucalyptus (Zafarullah, Anwar, & Anwar, 2011) |

In this research, we study three popular STaaS that have not been examined in the literature, namely: SpiderOak, JustCloud, and pCloud.

Users of SpiderOak, JustCloud and pCloud may download, upload, and access their data using web-browser and client application (e.g. mobile app). The SpiderOak client application also enables users to create, schedule, and restore backups and share files via password-protected links. In addition, the Hive feature of the SpiderOak client application allows contents to be synced across all devices linked to the user's account. SpiderOak Zero-knowledge service reportedly encrypts all data stored on their servers. Such a service would be attractive to cybercriminals and individual users who want to ensure that their stored data will not be surrendered to law enforcement agencies by the cloud service provider.

JustCloud allows users to synchronize, create backup, and share files. In file synchronization, a folder is created on the user's desktop after JustCloud's client application is installed, which enables the user to access data stored on all devices where the client application has been installed.

Users may store, sync, and share their files using pCloud. A unique feature of this CSSP is the upload link. User can have files uploaded on their space by those who have access to the upload link. This CSSP is also capable of making backup from other services including Dropbox, Facebook, Instagram, and Picasa. Table 2 briefly compares the features of aforementioned services.

Table 2: CSSPs Comparison Table

|  |  | SpiderOak | JustCloud | pCloud |
|---|---|---|---|---|
| Operating Systems Support | Windows | Y | Y | Y |
|  | Linux | Y | Y | Y |
|  | Mac OSX | Y | Y | Y |
|  | iOS | Y | Y | Y |
|  | Windows Phone | N | Y | N |
|  | Android | Y | Y | Y |
|  | BlackBerry | N | Y | N |
| Storage (by free) |  | 2GB | 15MB[1] | 20GB |
| Maximum paid storage |  | 100GB | Not Limited | 1TB |
| Backup |  | Y | Y | N |
| Sync |  | Y | Y | Y |
| Encryption |  | Y | Y | Y |
| Sharing |  | Y | Y | Y |

In this paper, we answer the following questions:

(1) What artifacts of forensic interest can be recovered from the Random Access Memory (RAM) and the Hard Disk (HDD) of a Windows device after using SpiderOak, JustCloud, and pCloud services via Internet Explorer (IE), Firefox (Fx), and Google Chrome (GC) browsers?
(2) What artifacts of forensic interest can be recovered from the RAM and HDD of a Windows device after using SpiderOak, JustCloud, and pCloud services via the respective client Windows applications?
(3) What artifacts of forensic interest can be recovered from the internal memory and internal storage of an iPhone device after using SpiderOak, JustCloud, and pCloud services via the respective iOS applications?
(4) Whether the contents or the metadata of the investigated files change during the process of uploading and downloading, and whether the timestamp information of the downloaded files is reliable?

---

[1] JustCloud free trial expires after 14 days.

Research will be conducted on Windows 8.1 and iOS devices. At the time of research, Windows 8.1 is the latest version of Microsoft desktop OS and iOS is one of the most popular mobile platforms (Net Applications, 2014a, 2014b).

The rest of the paper is organized as follows: In Section 2, we describe the forensic framework used to guide the research and the experiment setup. The findings from the analysis of SpiderOak, JustCloud, and pCloud are presented in sections 3, 4, and 5 respectively. Finally, the last section concludes this paper.

## 2. Research methodology

### 2.1 Cloud forensic framework

When conducting a forensic investigation, the investigator should adopt best practices such as those of the Association of Chief Police Officers (ACPO). The ACPO specifies four principles for collecting and examining digital evidence (Williams, 2012):

- Principle 1: Data which may subsequently be relied upon in a court of law should not be changed.
- Principle 2: In the event that a person needs to access the original data, that person must be suitably qualified and is able to justify and explain the implications of the actions.
- Principle 3: An appropriate auditing and record-keeping processes should be in place which would ensure that an independent third party would be able to examine the recorded processes and achieve the same result.
- Principle 4: The investigating officer needs to ensure that the law and these principles are adhered to.

It is also common practice that a forensic framework be used to guide the investigation. In the context of our paper, we adopt the cloud forensic framework introduced by Martini and Choo (Martini & Choo, 2012). This is, perhaps, the first digital forensic framework designed to conduct both client and server investigations of cloud services. The framework has also been validated by the authors using ownCloud (Martini & Choo, 2013), (Martini & Choo, 2014b), (Martini & Choo, 2014c), and by Thethi and Keane (2014) on EC2 cloud. There are four stages in this framework, namely: evidence source identification and preservation, collection, examination and analysis, and reporting and presentation for collecting digital evidence from the cloud environment.

1. *Evidence Source Identification, Collection and Preservation*. In this phase, potential sources of relevant data are identified. Any device capable of connecting to STaaS, either via a browser or a client application, is considered a potential source of evidence. In this phase, the investigator should also ensure that ACPO principles are adhered to, wherever possible. During collection of evidence from storage media, particularly media belonging to external parties, the investigator should also ensure that relevant laws and regulations are followed (Kent, Chevalier, Grance, & Dang, 2006).
In this research, the .vmem and vmdk files of each virtual machine (VM) were collected with extension E0 using AccessData FTK Imager. The former was cloned while the VM was running, whereas the latter was duplicated after the

VM was shut down. The logs of Wireshark, recording communications between VMs and the respective STaaS were also acquired at this stage. The MD5 hash checksum of collected evidence files were documented.
2. *Examination and analysis*. Information from acquired data is extracted in this phase. Methods to circumvent or bypass protection mechanism on the devices may be used to examine and analyze information collected and preserved from the previous phase (e.g. use of tools to brute-force password-protected data). During this phase, findings should also be reviewed with information or intelligence drawn from other sources and investigations (e.g. see metadata analysis described in Sections 3 to 5; and (D Quick & Choo, 2014) before a conclusion is drawn.
3. *Presentation*. In the last phase, findings are documented for presentation in a court of law (Kent et al., 2006).

## 2.2 Experiment setup

### 2.2.1 Windows

The Windows-based experiments were implemented on VMs using VMware Player 6.0.1. The following files with forensic value were collected and examined:

- .vmem file: A paging file that includes the backup of the VM's main memory (VMware Inc., n.d.).
- .vmdk file: A virtual disk file that stores the contents of the VM's hard disk drive (VMware Inc., n.d.).

Windows 8.1 Build 9600 along with IE 11.0.9600.16384, Fx 33.0.2, and GC 38.0.2125.111 m were installed on a VM with 25 GB hard disk and 1 GB memory. In this research, 14 files from a dd image file developed by Carrier (2004) were used as the dataset (see Table 3). The original dd image file was mounted on the VM using OSFMount 1.5.

Table 3. Files used in the Windows-based experiment

| Name | MD5 | Note |
|---|---|---|
| file1.jpg | 75b8d00568815a36c3809b46fc84ba6d | A JPEG file with JPEG extension |
| file2.dat | de5d83153339931371719f4e5c924eba | A JPEG file with non-JPEG extension |
| file3.jpg | 1ba4e91591f0541eda255ee26f7533bc | A non-image file with JPEG extension |
| file4.jpg | c8de721102617158e8492121bdad3711 | A non-image file with JPEG header |
| file5.rtf | 86f14fc525648c39d878829f288c0543 | A file with 0xffd8 signature value in multiple locations of the file. |
| file8.zip | d41b56e0a9f84eb2825e73c24cedd963 | A ZIP file with ZIP extension containing a JPEG file named file8.jpg. |
| file8.jpg | f9956284a89156ef6967b49eced9d1b1 | A JPEG file inside of the ZIP file |
| file9.boo | 73c3029066aee9416a5aeb98a5c55321 | A ZIP file with non-ZIP extension containing a JPEG file named file9.jpg |
| file9.jpg | c5a6917669c77d20f30ecb39d389eb7d | A JPEG file inside the ZIP file |
| file10.tar.gz | d4f8cf643141f0c2911c539750e18ef2 | A gzipped tar file containing a JPEG file named file10.jpg |
| file10.jpg | c476a66ccdc2796b4f6f8e27273dd788 | A JPEG file inside the gzipped tar file |
| file11.dat | f407ab92da959c7ab03292cfe596a99d | A JPEG file with dat extension. |
| file12.doc | 61c0b55639e52d1ce82aba834ada2bab | A Word document with a JPEG file inside it |
| file13.dll:here | 9b787e63e3b64562730c5aecaab1e1f8 | A JPEG file within an alternate data streams (ADS) |

From the base VM, the following snapshots were created:

- VM-W1: A SpiderOak account was created using IE.

- VM-W2: A SpiderOak account was created on Fx.
- VM-W3: A SpiderOak account was created using GC.
- VM-W4: A JustCloud account was created using IE.
- VM-W5: A JustCloud account was created on Fx.
- VM-W6: A JustCloud account was created using GC.
- VM-W7: A pCloud account was created on IE.
- VM-W8: A pCloud account was created on Fx in this VM.
- VM-W9: A pCloud account was created using GC.
- VM-W10: The SpiderOak's client application was installed. Sample files were also uploaded and downloaded using the client application.
- VM-W11: The JustCloud's client application was installed, and a series of uploading and downloading of sample files using the client application was undertaken.
- VM-W12: The pCloud's client application was installed.
- VM-W16: A series of downloading of sample files using SpiderOak with IE.
- VM-W17: Sample file were downloaded using SpiderOak via Fx.
- VM-W18: A series of downloading of sample files using SpiderOak with GC.
- VM-W19: A series of downloading of sample files using JustCloud with IE.
- VM-W20: A series of downloading of sample files using JustCloud with Fx.
- VM-W21: A series of downloading of sample files using JustCloud with GC.
- VM-W22: A series of downloading of sample files using pCloud with IE.
- VM-W23: A series of downloading of sample file using pCloud with Fx.
- VM-W24: A series of downloading of sample files using pCloud with GC.

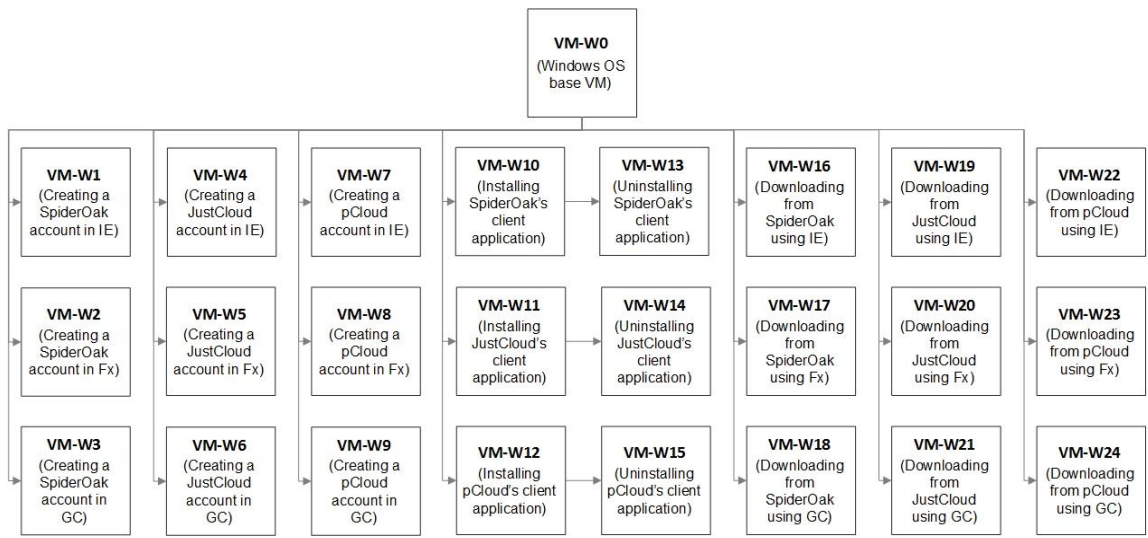

Figure 1. Overview of Windows-based Experiments

Each of the STaaS accounts created was linked to a specific email address created for this research (see Table 4).

Table 4. Research email accounts

| Email Address | First Name | Last Name |
| --- | --- | --- |
| csforensics_1@yahoo.com | Johnny | Appleseed |
| csforensics_2@ yahoo.com | Mary | Major |
| csforensics_3@ yahoo.com | Richard | Miles |

SpiderOak 5.1.8, JustCloud 1.4.0.28, and pCloud 1.3.2 were respectively installed on VM-W10, VM-W11, and VM-W12 and sample files were uploaded and downloaded using the respective client applications.

VM-W10, VM-W11, and VM-W12 were shut down and cloned respectively to VM-W13, VM-W14, and VM-W15. They were used for the investigation of data remnants after uninstallation of the client applications. On the last nine VMs (i.e. VM-W15 to VM-W24), the sample files were uploaded and downloaded using the respective browsers. On VM-W10, VM-W-11 and VM-W16 to VM-W24, MD5 checksum of the downloaded files were recorded and compared with the original ones.

In all our experiments, we configured the browsers and client applications to decline storing the passwords. In additional, prior to cloning each .vmem file, the corresponding browser was closed without the user signing out. The base VM was configured to create only one .vmdk file, and other VMs were generated by taking snapshots of the base .vmdk file. VMware Virtual Disk Manager Utility was used to merge the base .vmdk file with the snapshot files in order to make creating of the image file using FTK Imager possible.

*2.2.2 iOS*

In the iOS-based experiments, SpiderOak 3.1.1, JustCloud 1.3.11, and pCloud 1.11.15 apps were installed on an iPhone 5S device with iOS 8.1.1.

The following files were collected for further analysis on a personal computer (PC):

- Memory (.vmem file)
- Network logs collected by Wireshark (.PCAP files)
- Browsers' files
    - IE: %LocalAppData%\Microsoft\Windows\WebCache, %LocalAppData%\Microsoft\Internet Explorer\, and %LocalAppData%\Packages\windows_ie_ac_001\
    - Fx: %LocalAppData%\Mozilla\Firefox\Profiles\[Random Name].default\ and %AppData%\Mozilla\Firefox\Profiles\[Random Name].default\
    - GC: %LocalAppData%\Google\Chrome\User Data\Default
- Master File Table (MFT) with $MFT file name and located in %SystemDrive%
- NTFS log ($LogFile) located in %SystemDrive%
- Prefetch folder located in %SystemRoot%\Prefetch\
- Registry
- Paging file with pagefile.sys filename in %SystemDrive%
- Swap file named swapfile.sys in %SystemDrive%
- Windows events located in %SystemRoot%\System32\winevt\Logs\
- Link files
- Thumbcache files located in %LocalAppData%\Microsoft\Windows\Explorer\
- Unallocated space

The above collected files were analysed using following tools:

- Wireshark 1.12.1 was used to filter network traffic to detect IPs and ports used by the respective STaaS web-portal and client application.

- Autopsy Version 3.1.1 was used to conduct keyword search within E0 files linked to VMs' HDDs, and analyze browser cookies, browser histories and filtered files of E0 images.
- ESEDatabaseView v1.23 was used to analyse tables located in IE database. More specifically, for IE versions 10 and 11, the browser database is stored in an Extensible Storage Engine (ESE) file named WebCacheV01.dat (Malmström & Teveldal, 2013). The tables to be examined are contained in the ESE file located in %LocalAppData%\Microsoft\Windows\WebCache\.
- SQLite Manger was used to extract sqlite files of Fx and GC.
- DCode Date (Wilson, n.d.) was used for converting dates and times from hex, epoch, and WebKit formats to human readable format.
- Thumbcache Viewer 1.0.2.7 was used to extract thumbcache files.
- FTK Imager was used to analyze other files of interest.
- iExplorer was used for mounting and browsing iPhone backups.

Files from each of the E0 files that containing matching keywords were identified and filtered for further analysis – see Table 5.

Table 5. Keyword search items

| | VM-W1 to VM-W3 | VM-W4 to VM-W6 | VM-W7 to VM-W9 | VM-W10 | VM-W11 | VM-W12 | VM-W13 | VM-W14 | VM-W15 | VM-W16 to VM-W18 | VM-W19 to VM-W21 | VM-W22 to VM-W24 | The iPhone Device |
|---|---|---|---|---|---|---|---|---|---|---|---|---|---|
| Account email addresses | Y | Y | Y | Y | Y | Y | Y | Y | Y | Y | Y | Y | Y |
| Account names[2] | Y | Y | N/A | Y | Y | N/A | Y | Y | N/A | Y | Y | N/A | Y |
| Accounts ID[3] | Y | N/A | N/A | Y | NA | NA | Y | N/A | N/A | Y | N/A | N/A | Y |
| Account passwords | Y | Y | Y | Y | Y | Y | Y | Y | Y | Y | Y | Y | Y |
| CSSP names | Y | Y | Y | Y | Y | Y | Y | Y | Y | Y | Y | Y | N |
| Client applications' filenames | N | N | N | Y | Y | Y | Y | Y | Y | Y | Y | Y | N |
| Sample file names | N | N | N | Y | Y | Y | Y | Y | Y | Y | Y | Y | Y |
| Shared URLs | N | N | N | Y | N/A | N | Y | N/A | N | Y | N | Y | N |
| Share ID/Name | N | N/A | N/A | Y | N/A | Y | Y | N/A | Y | Y | N/A | Y | N |
| RoomKey | Y | N/A | N/A | Y | N/A | N/A | Y | N/A | N/A | Y | N/A | N/A | N |
| Shared files | N | N | N | Y | N/A | N | Y | Y | N | Y | Y | N | N |
| Shared URL password | N | N | N | Y | N/A | N/A | Y | N/A | N/A | Y | N/A | N/A | N |
| Invitation message | N/A | N/A | N | N/A | N/A | Y | N/A | N/A | Y | N/A | Y | Y | N |
| Synchronization name | N | N | N | N | Y | N/A | N | Y | N/A | Y | Y | N/A | N |
| Downloading path | N | N | N | Y | Y | Y | Y | Y | Y | Y | Y | Y | N/A |
| Devices' names | N | N | N | Y | Y | Y | Y | Y | Y | Y | Y | Y | N |
| Integrity check | N | N | N | Y | Y | N/A | N | N | N | Y | Y | Y | Y |

The iPhone backups were also searched for user email address, name, IDs, and password.

---

[2] pCloud registration process does not ask for the name of the account.

[3] Accounts of JustCloud and pCloud do not have ID.

## 3 Findings: SpiderOak

### *3.1 Observations: SpiderOak's account created using the respective browsers*

We were able to recover various information associated with the creation of the SpiderOak's account using the respective browsers – see Tables 6 to 8.

Table 6: Recovered artefacts associated with the creation of the SpiderOak's account using IE

| Location | Recovered artefacts |
|---|---|
| Network traffic | IP address 173.223.11.89 on port 80<br>38.121.104.79 and 38.121.104.80 on port 443 |
| Memory | Word "spideroak", the email address, the ID, and the name of the created account |
| Browser related files | The URL of SpiderOak visited during the creating of the account, in addition to the corresponding timestamp information of creation and access |
| Registry | URL of browsed webpages in HKEY_CURRENT_USER\Software\Microsoft\Internet Explorer\TypedURLs<br>The Timestamp information in HKEY_CURRENT_USER\Software\Microsoft\Internet Explorer\TypedURLsTime. |
| Paging | "spideroak.com" word |

Table 7: Collected artefacts with regard to the creation of the SpiderOak's account using Fx

| Location | Recovered artefacts |
|---|---|
| Network traffic | IP addresses 38.121.104.79 and 38.121.104.80 on port 443 |
| Memory | The "spideroak" word, email address, ID, and name of the created account |
| Browser related files | The email address used in signing up process (see Figure 2)<br>The date and the UTC time of visiting |
| Unallocated space | The email address, the ID and the full name of the registered account (see Figure 3) |
| Other files | SpiderOak's website was observed in a file located in %AppData%\Microsoft\Windows\Recent\CustomDestinations |

Table 8: Recovered artefacts associated with the creation of the SpiderOak's account using GC

| Location | Recovered artefacts |
|---|---|
| Network traffic | 38.121.104.79 and 38.121.104.80 on port 443 |
| Memory | "spideroak" word, the email address, the ID, the name, and the password of the created account (see Figure 4) |
| Browser related files | Information about visited spideroak.com with the date and the UTC time of the last visit and "spideroak.com" word, the email address, the full name, and the ID of the created account |
| Unallocated space | The email address of the created account |
| Other files | SpiderOak's website was seen in a file in %AppData%\Microsoft\Windows\Recent\CustomDestinations |

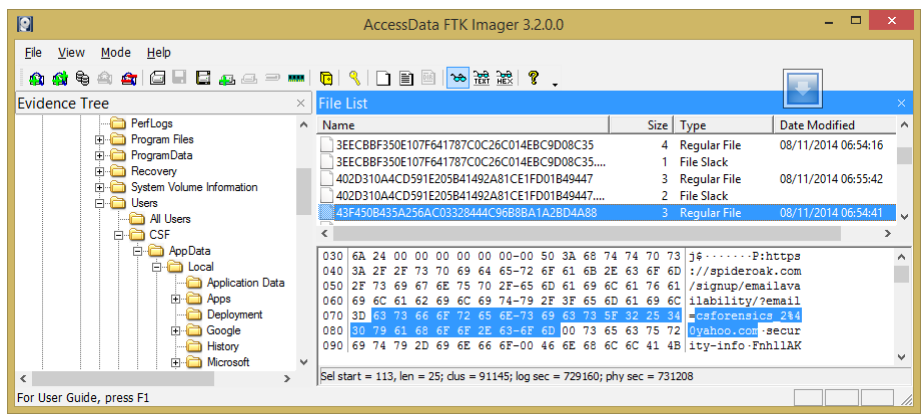

Figure 2. The email address of the created account

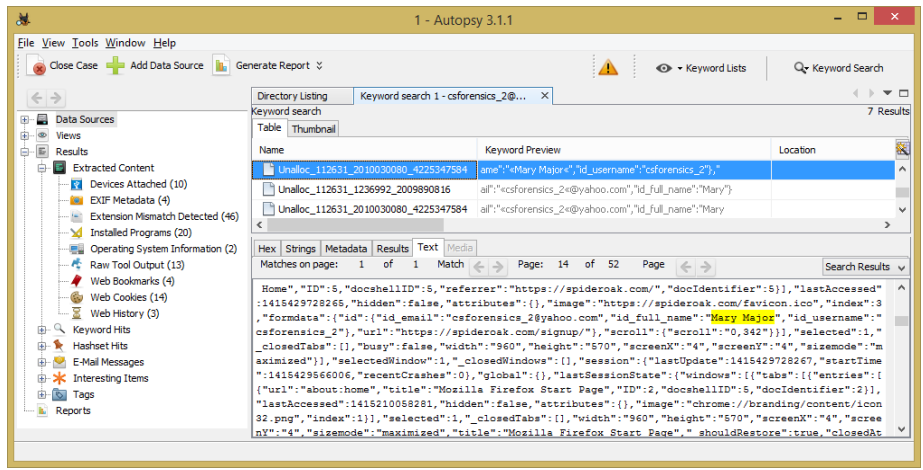

Figure 3. The email address, the ID, and the full name of the created account

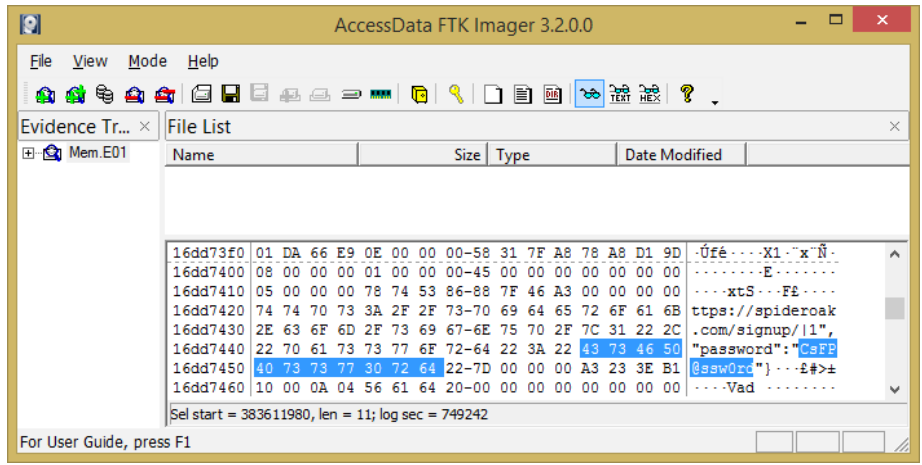

Figure 4. Recovered password of the created account

### 3.2 *Observations: SpiderOak's Application Program*

*Network traffics*. Communications between the VM and SpiderOak's servers were made with IP addresses 38.121.104.79, 38.121.104.89, and 38.121.104.90 on port 443.
*Memory*. SpiderOak, the username, the email address, and the password of the created the account along with the share ID (with the date and the time of creation), the shared URL and its password, the name of the shared folder, the name of the downloaded files, the name of the created sync and its folder on the host and on the target device, all were obtained from the collected vmem file (see Figure 5 and Figure 6).

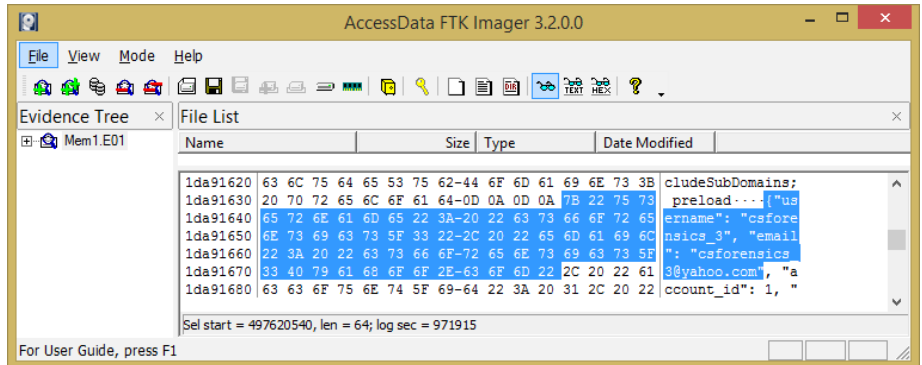

Figure 5. The username and the email address of SpiderOak's account

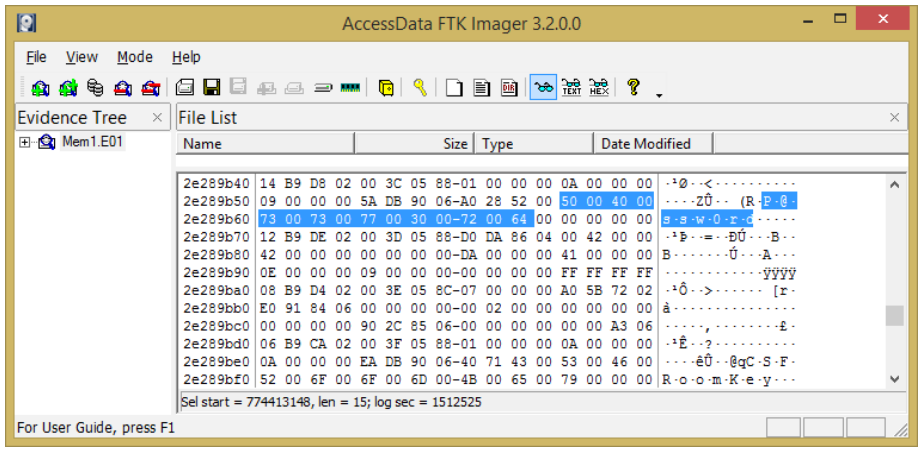

Figure 6. The password of the created shared URL in VM-W10's vmem file

*Client application files.* Client information such as the username, the share ID, RoomKey, the name of the shared folder, the name of the downloaded files, the name of the created sync and its folder on the host and on the target device as well as the date and the time of launched backup and synchronization were seen in %AppData%\SpiderOak\spider_###############.log. In the log file, an IP address was found which was appeared to be used for broadcasting on ports 21327 and 21328. The three first octal were the same as the device's IP address. In the oak_###############.log, similar events regarding the application along with the date and the time of occurrence were found. In %AppData%\SpiderOak\config.txt, the Hive path was observed. Files in %AppData%\SpiderOak\Sync contained the details of created syncs including the date and the time of their processing and source and destination paths, all, in SQLite format.

*MFT.* The sync name, the downloading folder name, and the installation filename were identified in the MFT$.

*Registry.* Installation filename and its storing path were found in the following keys:

- HKEY_CURRENT_USER\Software\Microsoft\Windows NT\CurrentVersion\AppCompatFlags\Compatibility Assistant\Store
- HKEY_USERS\S-1-5-21-1335463704-3291414260-3134846049-1001\Software\Microsoft\Windows NT\CurrentVersion\AppCompatFlags\Compatibility Assistant\Store

Moreover, SpiderOak's client application created inbound rules on both TCP and UDP protocols in Windows Firewall. The related keys' paths are as follow:

- [HKEY_LOCAL_MACHINE\SYSTEM\ControlSet001\Services\SharedAccess\Parameters\FirewallPolicy\FirewallRules]
- [HKEY_LOCAL_MACHINE\SYSTEM\CurrentControlSet\Services\SharedAccess\Parameters\FirewallPolicy\FirewallRules]

*Prefetch.* Installation filename was observed in Prefetch folder.
*Paging file.* The installation filename and the name of the downloaded files and their storing path were located in pagefile.sys.
*Windows events.* In the Application node, some records linked to the client application installation were located.

*Link files*. Two lnk files related to SpiderOak were located in %ProgramData%\Microsoft\Windows\Start Menu\Programs\SpiderOak\.
*Thumbcache*. A thumbnail of file1.jpg was located in the Thumbcache folder.
*Unallocated space*. The name of the downloaded files, the installation filename, and spideroak.exe string were located in the unallocated space.
*Other files*. We located some information concerning the downloading folder and the name of the downloaded files in %ProgramData%\Microsoft\Search\Data\Applications\Windows\GatherLogs\SystemIndex\SystemIndex.2.gthr and %LocalAppData%\Microsoft\Windows\Caches\{AFBF9F1A-8EE8-4C77-AF34-C647E37CA0D9}.1.ver0x0000000000000004.db, the Windows Indexing System and Windows File Caching respectively. The downloading folder name was also observed in NTUSER.DAT and ntuser.dat.LOG1, located in %UserProfile%, and the installation filename was observed in %SystemRoot%\AppCompat\Programs\Amcache.hve and %SystemRoot%\System32\config\SOFTWARE.LOG2. Also in %AppData%\SpiderOak\fs_queue.db, synchronized files with the date and the UTC time were identified.

Spideroak.exe was observed in the following files:

- %SystemRoot%\System32\config\SYSTEM and SYSTEM.LOG1
- %SystemRoot%\System32\config\
- %SystemRoot%\System32\wdi\LogFiles\WdiContextLog.etl.002
- %SystemRoot%\Temp\AA9B36AA-758C-4EB0-94A6-C5AB9F4CEC2A\CompatProvider.dll
- %LocalAppData%\Local\Microsoft\Windows\appsFolder.itemdata-ms.bak
- %LocalAppData%\Local\Microsoft\Windows\appsFolder.itemdata-ms
- %LocalAppData%\Local\Microsoft\Windows\Explorer\TileCacheLogo-1736031_100.dat
- %LocalAppData%\Local\Microsoft\Windows\Caches/{3DA71D5A-20CC-432F-A115-DFE92379E91F}.1.ver0x000000000000000b.db
- %UserProfile%\ntuser.dat.LOG2
- %SystemRoot%\AppCompat\Programs\Amcache.hve.LOG1
- %SystemRoot%\Installer\24c05.msi

Based on the collected MD5 checksums, apart from file13.dll, the contents of other sample files were preserved by the CSSP. We were not able to restore ADS from file13.dll. Metadata of file12.doc such as Authors value remained the same as the original file. It was also evident that the Modified value of downloaded files did not alter. However, we observed that the values of Created and Accessed fields were changed to the date and the time of downloading (see Figure 7).

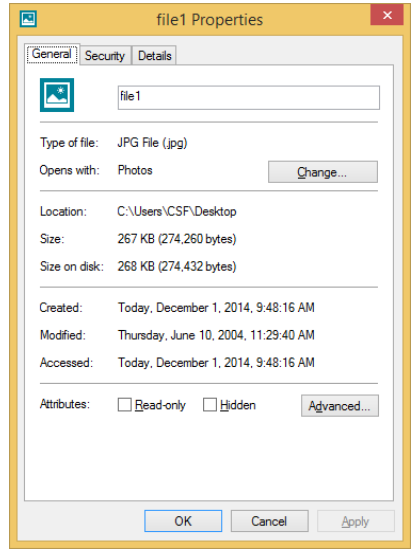

Figure 7. General tab of file1.jpg after being downloaded by SpiderOak's client application

## 3.3 Observations: Uninstalling SpiderOak's Application Program

*Memory*. Although the uninstall wizard asked for restart, restarting did not occur. Aside from the password of the shared URL, the result of memory analysis is similar to that reported in Section 3.2 (see Figure 8).

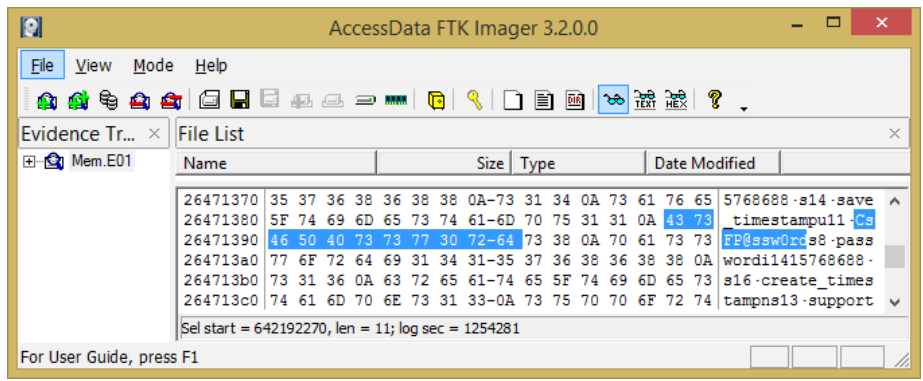

Figure 8. Account's password in memory of VM-W13

*Client application files*. After uninstalling SpiderOak on VM-W13, we observed that the Hive folder and its contents remained. We were also able to locate the files located in %AppData%\SpiderOak\ previously discussed in Section 3.2.
*MFT and paging file*. Both MFT and paging file contained spideroak word, the sync name, and downloading folder name.
*NTFS log*. "spideroak.exe" was observed in $LogFile.
*Registry*. The following keys pointed to spideroak.exe after uninstalling:

- [HKEY_CLASSES_ROOT\*\shellex\ContextMenuHandlers\SpiderOak]
- [HKEY_CLASSES_ROOT\Directory\background\shellex\ContextMenuHandlers\SpiderOak]

- [HKEY_CLASSES_ROOT\Directory\shellex\ContextMenuHandlers\SpiderOak]
- [HKEY_CLASSES_ROOT\Local Settings\Software\Microsoft\Windows\Shell\MuiCache]
- [HKEY_CURRENT_USER\Software\Classes\Local Settings\Software\Microsoft\Windows\Shell\MuiCache]
- [HKEY_CURRENT_USER\Software\Microsoft\Windows\CurrentVersion\Run]
- [HKEY_CURRENT_USER\Software\Microsoft\Windows\CurrentVersion\UFH\SHC]
- [HKEY_LOCAL_MACHINE\SOFTWARE\Classes\*\shellex\ContextMenuHandlers\SpiderOak]
- [HKEY_LOCAL_MACHINE\SOFTWARE\Classes\Directory\background\shellex\ContextMenuHandlers\SpiderOak]
- [HKEY_LOCAL_MACHINE\SOFTWARE\Classes\Directory\shellex\ContextMenuHandlers\SpiderOak]
- [HKEY_LOCAL_MACHINE\SYSTEM\ControlSet001\Services\SharedAccess\Parameters\FirewallPolicy\FirewallRules]
- [HKEY_LOCAL_MACHINE\SYSTEM\CurrentControlSet\Services\SharedAccess\Parameters\FirewallPolicy\FirewallRules]
- [HKEY_USERS\S-1-5-21-1335463704-3291414260-3134846049-1001\Software\Classes\Local Settings\Software\Microsoft\Windows\Shell\MuiCache]
- [HKEY_USERS\S-1-5-21-1335463704-3291414260-3134846049-1001\Software\Microsoft\Windows\CurrentVersion\Run]
- [HKEY_USERS\S-1-5-21-1335463704-3291414260-3134846049-1001\Software\Microsoft\Windows\CurrentVersion\UFH\SHC]
- [HKEY_USERS\S-1-5-21-1335463704-3291414260-3134846049-1001_Classes\Local Settings\Software\Microsoft\Windows\Shell\MuiCache]

*Prefetch folder.* "spideroak.exe" was observed in AgRobust.db, Layout.ini, Trace2.fx, RUNDLL32.EXE-125D4518.pf, RUNDLL32.EXE-F61C91E8.pf, and SPIDEROAK.EXE-CA2F0025.pf.

*Windows events.* In the Application node, Microsoft-Windows-Diagnostics-Performance%4Operational.evtx, Microsoft-Windows-Shell-Core%4Operational.evtx, and Microsoft-Windows-Windows Firewall With Advanced Security%4Firewall.evtx, we located records indicating the uninstalling of the client application. In addition, spideroak.exe was observed in WdiContextLog.etl.001, WdiContextLog.etl.002, and WdiContextLog.etl.003 which were located in %SystemRoot%\System32\wdi\LogFiles\.

*Link file.* %ProgramData%\Microsoft\Windows\Start Menu\Programs\SpiderOak\SpiderOak.lnk was linked to spideroak.exe.

*Unallocated space.* The spideroak word, the sync name, the sync folder name on the target machine, and downloading folder name were located in the unallocated space.

*Other files.* We located information indicating the name of the downloading folder in %ProgramData%\Microsoft\Search\Data\Applications\Windows\GatherLogs\SystemIndex\SystemIndex.2.gthr and %LocalAppData%\Microsoft\Windows\Caches\{AFBF9F1A-8EE8-4C77-AF34-C647E37CA0D9}.1.ver0x0000000000000004.db. Downloading folder name was also seen in NTUSER.DAT, and ntuser.dat.LOG1. The latter was located in %UserProfile%. Several other files we located also contained the spideroak.exe word.

### *3.4 Observations: Downloading from SpiderOak using the respective browsers*

We were able to collect information with forensic value associated with the downloading from SpiderOak using the respective browsers – see Tables 9 to 11.

Table 9: Recovered artefacts associated with downloading from SpiderOak using IE

| Location | Recovered artefacts |
|---|---|
| Network traffic | IP addresses 38.121.104.80, 80.157.17.91, and 204.79.197.200 on port 80 and 38.121.104.79 on port 443 |
| Memory | The username and the device name were visible in the URL (see Figure 9), shared title – the page was browsed -, and download folder name (SO Downloaded Files) |
| Browser logs | The URLs of SpiderOak that were visited along with the date and the UTC time of browsing, the downloading folder name and its path, the username stated in a URL, and the device name |
| MFT and NTFS log | The device name, the downloading folder name, the RoomKey, and the shared title (see Figure 10) |
| Registry | In addition to Typed URL, the following keys pointed to spideroak.com:<br><br>- HKEY_CLASSES_ROOT\Local Settings\Software\Microsoft\Windows\CurrentVersion\AppContainer\Storage\windows_ie_ac_001\Internet Explorer\DOMStorage\spideroak.com<br><br>- HKEY_CURRENT_USER\Software\Classes\Local Settings\Software\Microsoft\Windows\CurrentVersion\AppContainer\Storage\windows_ie_ac_001\Internet Explorer\DOMStorage\spideroak.com<br><br>- HKEY_USERS\S-1-5-21-1335463704-3291414260-3134846049-1001\Software\Classes\Local Settings\Software\Microsoft\Windows\CurrentVersion\AppContainer\Storage\windows_ie_ac_001\Internet Explorer\DOMStorage\spideroak.com<br><br>HKEY_USERS\S-1-5-21-1335463704-3291414260-3134846049-1001_Classes\Local Settings\Software\Microsoft\Windows\CurrentVersion\AppContainer\Storage\windows_ie_ac_001\Internet Explorer\DOMStorage\spideroak.com |
| Paging | The shared title |
| Unallocated Space | Contained the device name and the shared title |
| Other files | The shared title was seen in:<br>- %SystemRoot%\WinSxS\x86_microsoft-windows-twinui.resources_31bf3856ad364e35_6.3.9600.16384 |
|  | _en-us_c34c6f62bd49d58b\twinui.dll.mui<br>• %LocalAppData%\Packages\windows_ie_ac_001\AC\INetCache\XSEVJ9WJ\shares[1].json<br>• %SystemRoot%\System32\en-US\twinui.dll.mui<br>Downloading folder name was observed in:<br>• %UserProfile%\NTUSER.DAT and ntuser.dat.LOG1<br>• %SystemDrive%\$Extend/$UsnJrnl:$J<br>• %SystemRoot%\System32\wdi/LogFiles\BootCKCL.etl<br>• %LocalAppData%\Microsoft\Windows\UsrClass.dat and UsrClass.dat.LOG1<br>• %ProgramData%\Microsoft\Search\Data\Applications\Windows\GatherLogs\SystemIndex\SystemIndex.2.gthr<br>Two files, one pointing to the downloading folder name and the other linked to the downloaded folder were found in %AppData%\Microsoft\Windows\Recent\. |

Table 10: Recovered artefacts associated with downloading from SpiderOak using Fx

| Location | Recovered artefacts |
|---|---|
| Network traffic | IP addresses 38.121.104.80 on port 80 and 38.121.104.79 on port 443 |
| Memory | The username and the device name in plaintext located in URL, and the shared title |
| Browser logs | Visited SpiderOak's URLs with the date and the UTC time were observed in cookies.sqlite; and places.sqlite located in %AppData%\Mozilla\Firefox\Profiles\[Random Name].default\ (see Figures 10 and 11).<br>The name of the downloaded file<br>Both sessionstore.js and places.sqlite-wal were located in the same path contained the shared title and the username, and the shared title were located in several files in %LocalAppData%\Mozilla\Firefox\Profiles\[Random Name].default\cache2\entries\ and a file located in %AppData%\Microsoft\Windows\Recent\CustomDestinations\. The latter also included the username and the name of the downloaded file. The username was also found in %ProgramFiles%\Google\Update\1.3.25.5\goopdateres_is.dll.<br>As depicted in Figure 12, in places.sqlite and the moz_annos table, the name of the downloaded file was observed. |
| MFT | spideroak.com and the downloaded filename were seen in $MFT |
| NTFS log | The name of the downloaded file was seen in $LogFile |
| Link | The only related link file observed was %AppData%\Microsoft\Windows\Recent\e_.lnk, which referred to an archive of all sample files downloaded from SpiderOak |
| Unallocated space | spideroak.com, the device name, username were visible in the URL, and the downloaded filename were located in the unallocated space. |

| Other files | The name of the downloaded file was observed in the below files: |
| --- | --- |
| | • %LocalAppData%\Microsoft\Windows\WebCache\V01.log and WebCacheV01.dat |
| | • %UserProfile%\NTUSER.DAT and ntuser.dat.LOG1 |
| | • %SystemDrive%\$Extend\$UsnJrnl:$J |

Table 11: Recovered artefacts associated with downloading from SpiderOak using GC

| Location | Recovered artefacts |
| --- | --- |
| Network traffic | IP addresses 38.121.104.80 on port 80 and 38.121.104.79 on port 443 |
| Memory | The username and the device name were visible in the URL |
| Browser logs | History and Cookies files contained information about visited spideroak URLs with the date and the UTC time of browsing. In the History file, the name of the downloaded file and its associated device name were observed. History Provider Cache, Cookies-journal, Shortcuts, Top Sites, Top Sites-journal, Favicons, Favicons-journal, Preferences, Current Tabs, and Current Session were located in %LocalAppData%\Google\Chrome\User Data\Default, which pointed to spideroak.com. The username and the device name were also observed in several files located in %LocalAppData%\Google\Chrome\User Data\Default\Cache\. |
| MFT and NTFS log | "Spideroak.com" and the name of the downloaded file |
| Unallocated space | "spideroak" word |
| Other files | A file in %AppData%\Microsoft\Windows\Recent\CustomDestinations\ pointing to a URL containing the username in plaintext |

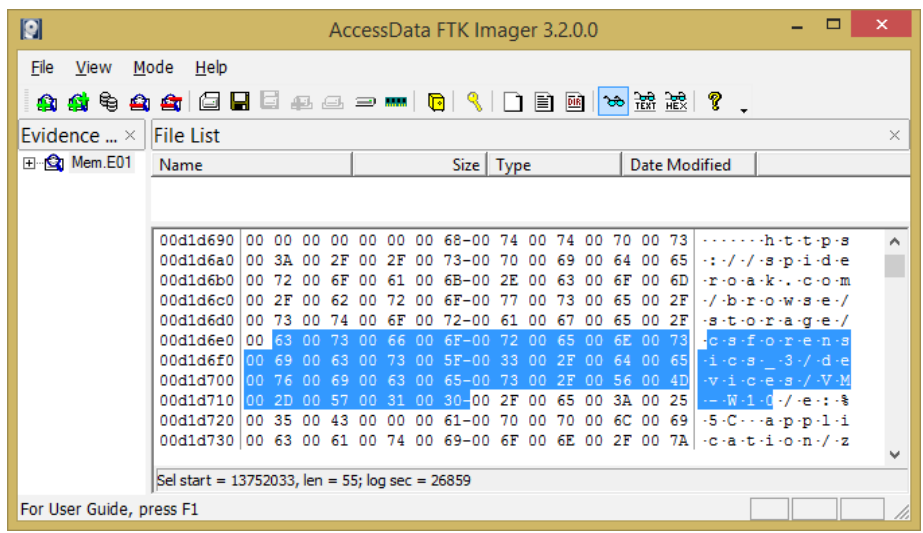

Figure 9. The username and the device name in the collected vmem file from VM-W16

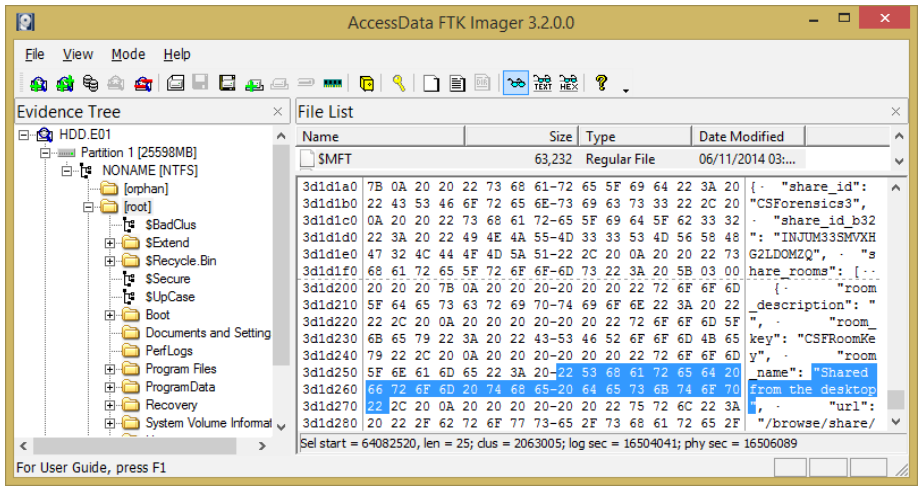

Figure 10. RoomKey and shared message in the MFT

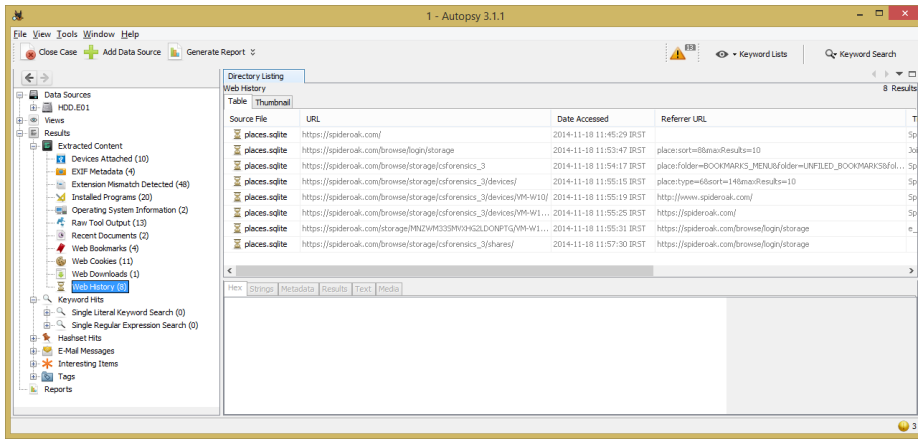

Figure 11. The username and the device name in the URLs

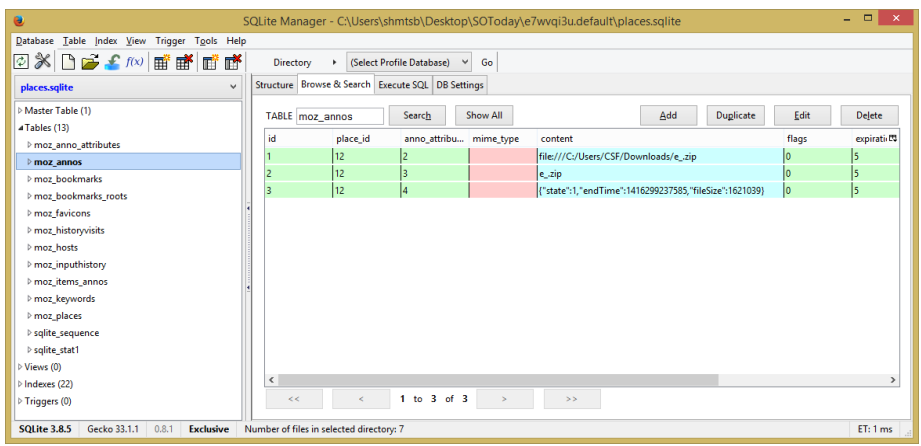

Figure 12. The username and the device name in places.sqlite

According to the recorded MD5 checksums, the contents of the downloaded files, with the exception of file13.dll, remained intact. We were not able to restore ADS from file13.dll. Metadata of file12.doc such as Authors value remained the same as the original file. Unlike the result presented in Section 3.2, in addition to Created and

Accessed fields, Modified value was also changed to the date and the time of downloading.

### 3.5 Observations: Browsing and downloading from SpiderOak's iOS app

As shown in Figure 13, a file named Cache.db-wal located in \Apps\SpiderOak\Library\Caches\com.spideroak.SpiderOak\nsurlcache contained valuable information regarding the registered account and created backups and syncs.

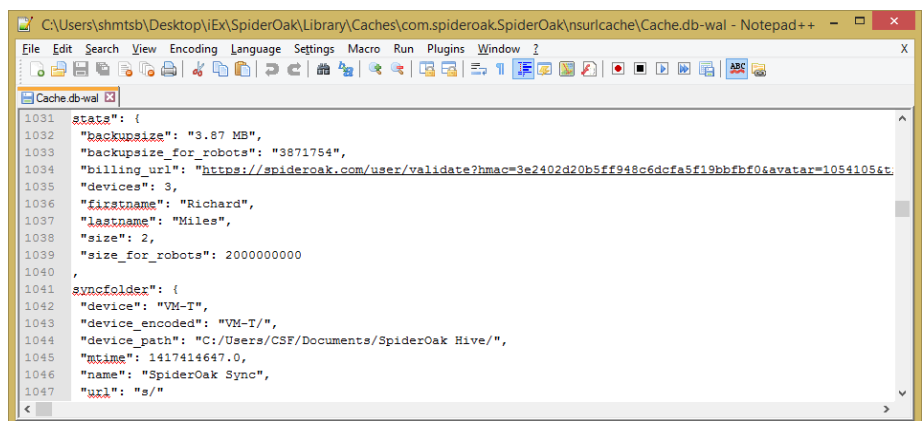

Figure 13. Cache.db-wal containing some details of the account and created backups and syncs

Downloaded files were located in \Apps\SpiderOak\tmp\.caches, and \Apps\SpiderOak\Library\Caches\com.spideroak.SpiderOak\nsurlcache\fsCachedData.
The dates and the times of Modified fields were identical to the date and the time of loading into the smartphone.
Apart from the file13.dll, the MD5 signatures of the sample files were the same as the original files. Modified date was pointing to the download time.

## 4 Findings: JustCloud

### 4.1 Observations: JustCloud's account created in using IE the respective browsers

Recovered information associated with the creation of the JustCloud's account using the respective browsers are presented in Tables 12 to 14.

Table 12: Recovered artefacts associated with the creation of the JustCloud's account using IE

| Location | Recovered artefacts |
|---|---|
| Network traffic | IP addresses 184.154.150.158, 37.252.162.208, 54.246.123.239, 74.125.230.154, 104.219.49.71, 94.31.29.154, 82.178.158.19, 74.125.230.132, 54.88.49.148, 217.163.21.35, 184.154.150.133, 173.194.67.95, 94.31.29.237, 82.178.158.18, 74.125.230.153, 23.251.138.168, and 10.10.34.34 on port 80. We also determined that some communications took place via IP address 184.154.150.158 on port 443. The email address used for creating the account was found in a packet with 184.154.150.158:80 destination. |
| Memory | The "justcloud" word, the email address and the name of the created |

| | account |
|---|---|
| Browser related files | The JustCloud URLs visited, along with the date and the UTC time of browsing were found in WebCacheV01.dat.<br>The full name and the email address of the created account along with "justcloud" word in %LocalAppData%\Microsoft\Internet Explorer\Recovery\Last Active\{056F2E5F-67E5-11E4-9717-000C29B56A39}.dat and %LocalAppData%\Packages\windows_ie_ac_001\AC\INetCache\XSEVJ9WJ\account[1].htm. |
| Registry | The URL of the visited webpages and the date and the UTC time of browsing respectively in TypedURLs and TypedURLsTime keys. |
| Unallocated space | The name and the email address of the created account |

Table 13: Recovered artefacts associated with the creation of the JustCloud's account using Fx

| Location | Recovered artefacts |
|---|---|
| Network traffics | IP addresses 184.154.150.158, 37.252.170.92, 54.75.236.238, 74.125.230.154, 94.31.29.154, 92.122.210.114, 74.125.136.95, 64.233.167.139, 54.164.48.8, 217.163.21.35, 104.219.49.71, 94.31.29.237, 74.125.230.141, 23.251.136.174, 184.154.150.133, and 10.10.34.34 on port 80. We also determined that some communications took place via IP address 184.154.150.158 on port 443. The email address used for creating the account was found in a packet with 184.154.150.158:80 destination. |
| Memory | The "justcloud" word and the email address of the created account were observed in the collected vmem file. |
| Browser related files | In %LocalAppData%\Mozilla\Firefox\Profiles\[Random Name].default\cache2\entries, the following files were indicating justcloud.com was browsed:<br>• 09ACE8E5F96EFE3AA5A0CE0B5A5563299A0FDA44<br>• 1810CD15A41C6F610D8C7E8C4D39CDACB00E337A<br>• 1BDE5F2D8D9F11EF0F335C5A9128DBE71382DA75<br>• 380C3CF91DFA451FBF740CA76A7DCB2BDBC0889C<br>• 3E077C83363C5F69ACAEACB5C1994E04464AC18F<br>• 446FF2398AF1F92765363BD3327D6D8CA4F77EC3<br>• 461025C8C3F848EBC71F9E28B9120ABF00A1E074<br>• 54872F69BC90AE690242ED63D0318EDA90D3557C<br>• 553B5944009C5918AF6AF5A09812A8B0D17E6FBF<br>• 5AF6B030BB839756C1D72CDB22A272BF2A618C22<br>• 6457C1BCEB28A521EE0ACC12400A002A8C1D1A8A<br>• 7280D1006F4EA8E782FF4988DDACC2FE45E223CE<br>• 7B44045EC0465D6CCFCB9E9FB691424E2E970466<br>• 7F74C3A115A0B08422A8409FC7494E00E9090F75<br>• 9252811649D77766E4FB7DC750D825FE68E4CF10<br>• 93B403780DF9FED67C2E7AE5EC879780BFAF41C1<br>• A721B51969F936D4EBAA8E4C6E3F99F8B703241B<br>• AE3FAB291B124D80634CA386411F72B43E1B300D<br>• C0EE83175D8130C265E78027748B480ABA944418 |

|  | - C83034755D57A645F6FF927E13522DCA8EC5E39A |
|  | - CDC7FBBA3E3BA42AFF33CD41F35C9EB8D63C0497 |
|  | - CF49CBF7AD56117C73D75F827975BA8D9CFDF67B |
|  | - D03FD20E97FDEE10EE8D20A4096B467ADB0B75B2 |
|  | - E84B0DEA1E343D9D0C39D74D7D8AED548046BD3C |
|  | - F0016B4F7C6CB097EA6FB493CB9F3070AA8ABA37 |
|  | - F588D60C1323FF25E996F26CD45405D82D11C36E |
|  | - F741938599B126A67F07C77106C86175D4F2DDBE |
|  | - FC83C509516A560D495BCDD9C0C0025E0BB16907 |
|  | Information regarding cookies and histories were found in cookies.sqlite, places.sqlite, and permissions.sqlite. The date and the UTC time of visiting were retrieved from the two first files. |
| Unallocated space | The full name and the email address of the created account |

Table 14: Recovered artefacts associated with the creation of the JustCloud's account using GC

| Location | Recovered artefacts |
| --- | --- |
| Network traffics | IP addresses 184.154.150.158, 192.168.74.139, 68.67.128.156, 173.194.112.77, 54.246.114.22, 104.219.49.71, 217.163.21.34, 94.31.29.154, 74.125.136.95, 208.71.121.1, 195.59.150.17, 173.194.112.68, 107.21.26.11, 94.31.29.237, 23.251.128.113, 207.46.194.10, 204.79.197.200, 184.154.150.133, and 10.10.34.34 on port 80. It was also determined that communications also took place via IP address 184.154.150.158 on port 443. |
| Memory | The "justcloud" word and the email address, the name, and the password of the created account |
| Browser files. | History and Cookies files contained information about visited JustCloud's URLs with the date and the UTC time of browsing. Shortcuts, Top Sites, Favicons, Current Tabs, and Current Session all located in %LocalAppData%\Google\Chrome\User Data\Default also pointed to justcloud.com. |

**4.2 *Observations: JustCloud's Application Program***

*Network traffics*. Communications between the client application and JustCloud's servers were made with IP addresses 54.231.64.4 and 184.154.150.133 on port 80.
*Memory*. The installation filename and its storing path, the name and the email address of the created account, the date and the time of account creation were observed in the memory.

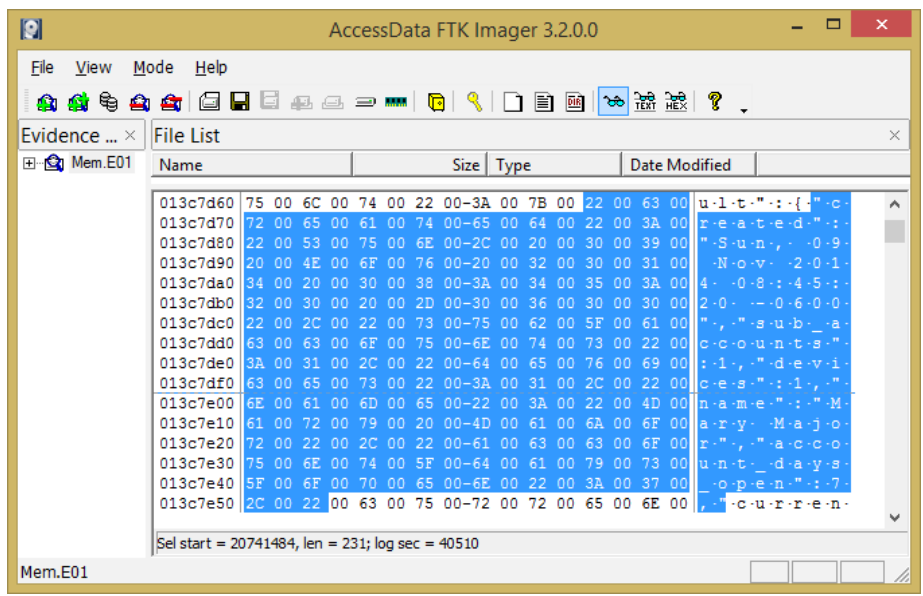

Figure 14. The name of the registered account along with account creation date and time in vmem file of VM-W11

*Client application files*. In %ProgramFiles%\JustCloud\Database, the following files of forensic interest were identified:

- mpcb_settings.db contained the email address and the computer name used to log in.
- mpcb_backup_conf.db included the name of the files that were browsed to be backed up.

  Also in %ProgramFiles%\JustCloud\log, the below key files were found:

- BACKUP.log contained a history of created backups along with the date and the time.
- LICENCE.log included the number of days that the account had been created and the name of the account.
- APPLICATION.log contained the date and the time of executing the program.
- DOWNLOADER.log included the details of a download failure occurred during downloading sample files.
- ONLINE_FOLDER.log contained the names of downloaded files.

*MFT, Pagefile.sys, and unallocated*. These contained the installation filename and JustCloud.exe.
*NTFS log*. The installation filename and JustCloud.exe were observed in $LogFile.
*Prefetch folder*. CONSENT.EXE-65F6206D.pf, JUSTCLOUD_SETUP.EXE-3C2F187F.pf, AgRobust.db, JUSTCLOUD.EXE-A4042D73.pf, RUNDLL32.EXE-125D4518.pf, and TASKENG.EXE-5BAF290C.pf located in %SystemRoot%\Prefetch\ included either the installation filename or JustCloud.exe. Layout.ini in the same path contained both mentioned files.
*Windows events*. An event regarding JustCloud's backup service was located in the System log.
*Link Files*. Four relevant lnk files were found:

- %AppData%\Microsoft/Windows/Start Menu/Programs/JustCloud/JustCloud.lnk
- %AppData%\Microsoft/Windows/Start Menu/Programs/Startup/JustCloud.lnk
- %UserProfile%\Desktop\JustCloud.lnk
- %UserProfile%\Desktop\Sync Folder.lnk

*Thumbcache files*. JustCloud's icon was located in the thumbcache files.
*Other files*. A task named LaunchApp was created in %SystemRoot%\System32\Tasks\ and scheduled to run JustCloud.exe /windowlaunch command. We observed the installation filename in the following files:

- %LocalAppData%\IconCache.db
- %ProgramFiles%\JustCloud\spf.dat
- %SystemDrive%\$Extend\$UsnJrnl:$J
- %SystemRoot%\AppCompat\Programs\Amcache.hve and Amcache.hve.LOG1
- %SystemRoot%\System32\config\RegBack\SYSTEM
- %SystemRoot%\System32\config\SYSTEM, SYSTEM.LOG1, and SYSTEM.LOG2
- %SystemRoot%\System32\sru\SRU.log and SRUDB.dat
- %SystemRoot%\System32\wdi\LogFiles\BootCKCL.etl
- %UserProfile%\NTUSER.DAT and ntuser.dat.LOG1

The below files contained "JustCloud.exe":

- %SystemDrive%\$Extend\$UsnJrnl:$J
- %LocalAppData%\IconCache.db
- %LocalAppData%\Microsoft\Windows\Caches\{3DA71D5A-20CC-432F-A115-DFE92379E91F}.1.ver0x000000000000000c.db
- %UserProfile%\NTUSER.DAT and ntuser.dat.LOG1
- %SystemRoot%\AppCompat\Programs\Amcache.hve and Amcache.hve.LOG1
- %SystemRoot%\System32\config\RegBack\SOFTWARE
- %SystemRoot%\System32\config\SOFTWARE, SOFTWARE.LOG1, SOFTWARE.LOG2, SYSTEM, and SYSTEM.LOG1.
- %SystemRoot%\System32\config\TxR\{b7bee95a-0b1a-11e3-93fc-90b11c043665}.TxR.0.regtrans-ms
- %SystemRoot%\System32\sru\SRU.log, SRU00008.log, and SRUDB.dat

From our experiments, we determined that JustCloud does not allow the uploading of files with an invalid extension and users attempting to do so will be presented with an error message – "*Failed: The remote server returned an error: (403) Forbidden.*". Recorded MD5 checksums showed that, the contents of sample files, with the exception of file13.dll, were preserved by this CSSP. Similar to the findings presented in Section 3.2, we were not able to restore ADS from file13.dll. Metadata of file12.doc were not altered. The values of Created, Modified, and Accessed fields were changed to the date and the time of downloading.

### *4.3 Observations: Uninstalling JustCloud's Application Program*

*Memory*. With the exception of not able to locate the email address of the account, the investigation result of the vmem file was the same as Section 4.2.

*Client application files.* Although the JustCloud folder was removed by the uninstalling process, they could be easily recovered using FTK Imager (see Figure 15).

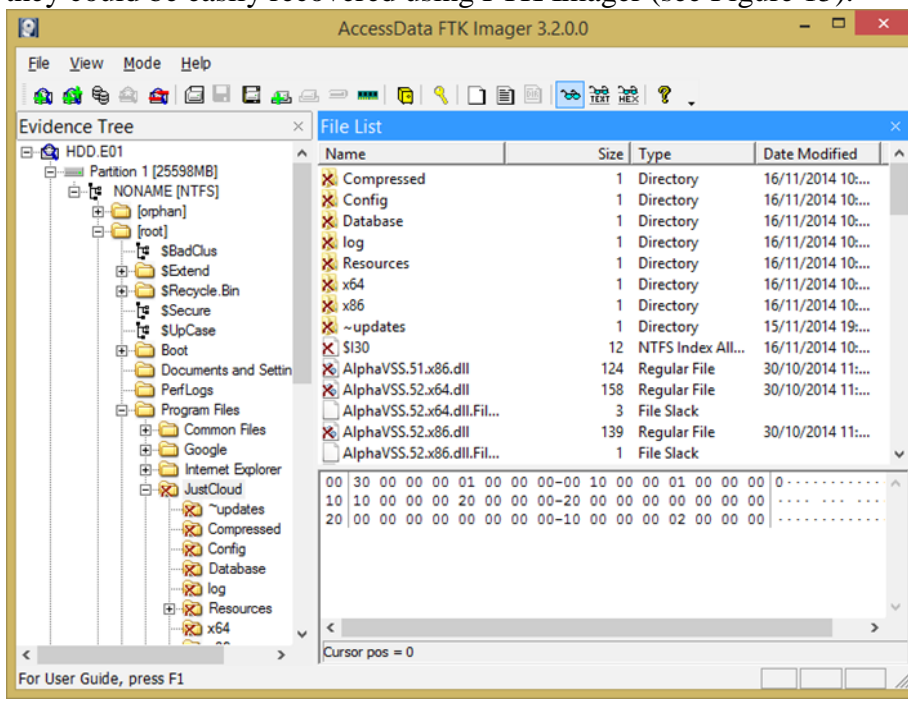

Figure 15. Deleted folders of JustCloud

*MFT and paging file.* The installation filename and justcloud.exe were observed in MFT$ and pagefile.sys.
*NTFS log.* The installation filename and justcloud.exe were observed in $LogFile.
*Registry.* The following keys pointed to JustCloud.exe after uninstallation:

- HKEY_CURRENT_USER\Software\Microsoft\Windows\CurrentVersion\UFH\SHC
- HKEY_CURRENT_USER\Software\Microsoft\Windows NT\CurrentVersion\AppCompatFlags\Compatibility Assistant\Store
- HKEY_LOCAL_MACHINE\SOFTWARE\Microsoft\Tracing\JustCloud_RASAPI32
- HKEY_LOCAL_MACHINE\SOFTWARE\Microsoft\Tracing\JustCloud_RASMANCS
- HKEY_USERS\S-1-5-21-1335463704-3291414260-3134846049-1001\Software\Microsoft\Windows\CurrentVersion\UFH\SHC
- HKEY_USERS\S-1-5-21-1335463704-3291414260-3134846049-1001\Software\Microsoft\Windows NT\CurrentVersion\AppCompatFlags\Compatibility Assistant\Store

*Prefetch.* In the Prefetch folder, CONSENT.EXE-65F6206D.pf, JUSTCLOUD_SETUP.EXE-3C2F187F.pf, and Layout.ini contained the installation filename. Also JustCloud.exe was included in AgRobust.db, BACKUPSTACK.EXE-831F34B6.pf, Layout.ini, TASKENG.EXE-5BAF290C.pf, RUNDLL32.EXE-125D4518.pf, and JUSTCLOUD.EXE-A4042D73.pf.
*Windows events.* A record in %SystemRoot%\System32\winevt\Logs\Microsoft-Windows-Diagnostics-Performance%4Operational.evtx pointed to justcloud.com.

*Unallocated*. The installation filename and the name of the account, backed up filenames with their paths, justcloud.com, and justcloud.exe were located in the unallocated space.

*Other files*. The scheduled task (LaunchApp) existed even after uninstallation. We observed the installation filename in the following files:

- %LocalAppData%\IconCache.db
- %ProgramFiles%\JustCloud\spf.dat
- %SystemDrive%\$Extend\$UsnJrnl:$J
- %SystemRoot%\AppCompat\Programs\Amcache.hve
- %SystemRoot%\System32\config\RegBack\SYSTEM
- %SystemRoot%\System32\config\SYSTEM and SYSTEM.LOG1
- %SystemRoot%\System32\sru\SRUDB.dat and SRUtmp.log
- %SystemRoot%\System32\wdi\LogFiles\BootCKCL.etl
- %UserProfile%\NTUSER.DAT and ntuser.dat.LOG1

    Also the below files were pointing to JustCloud.exe:

- %SystemRoot%\System32\wdi\{86432a0b-3c7d-4ddf-a89c-172faa90485d}\{4d3b40fb-cace-4558-a9ad-cbd7532ac6e2}\snapshot.etl
- %SystemRoot%\System32\wdi\{86432a0b-3c7d-4ddf-a89c-172faa90485d}\S-1-5-21-1335463704-3291414260-3134846049-1001_UserData.bin
- %SystemRoot%\System32\wdi\{86432a0b-3c7d-4ddf-a89c-172faa90485d}\{fbde4aa3-f5b4-4dc2-8312-ae0f4e1228c8}\snapshot.etl
- %SystemRoot%\System32\wdi\{86432a0b-3c7d-4ddf-a89c-172faa90485d}\{73b44f2d-609e-4476-8a08-9342afe75051}\snapshot.etl
- %LocalAppData%\IconCache.db
- %LocalAppData%\Microsoft\Windows\Caches\{3DA71D5A-20CC-432F-A115-DFE92379E91F}.1.ver0x000000000000000d.db and {3DA71D5A-20CC-432F-A115-DFE92379E91F}.1.ver0x000000000000000c.db
- %UserProfile%\NTUSER.DAT, ntuser.dat.LOG1, and ntuser.dat.LOG2
- %SystemRoot%\AppCompat\Programs\Amcache.hve and Amcache.hve.LOG1
- %SystemDrive%\$Extend\$UsnJrnl:$J
- %SystemRoot%\System32\config\RegBack\SOFTWARE and SYSTEM
- %SystemRoot%\System32\config\ SYSTEM, SYSTEM.LOG1, SOFTWARE.LOG1, SOFTWARE.LOG2 and SOFTWARE
- %SystemRoot%\System32\config\TxR\{b7bee95a-0b1a-11e3-93fc-90b11c043665}.TxR.0.regtrans-ms
- %SystemRoot%\System32\sru\SRUDB.dat and SRUtmp.log
- %SystemRoot%\System32\wdi\LogFiles\BootCKCL.etl
- %SystemRoot%\System32\wdi\LogFiles\StartupInfo\S-1-5-21-1335463704-3291414260-3134846049-1001_StartupInfo2.xml, S-1-5-21-1335463704-3291414260-3134846049-1001_StartupInfo5.xml, and S-1-5-21-1335463704-3291414260-3134846049-1001_StartupInfo1.xml
- %SystemRoot%\System32\wdi\LogFiles\ WdiContextLog.etl.001, WdiContextLog.etl.002, and WdiContextLog.etl.003

### *4.4 Observations: Downloading from JustCloud using the respective browsers*

We were able to collect information with forensic value associated with the downloading from JustCloud using the respective browsers – see Tables 15 to 17.

Table 15: Recovered artefacts associated with downloading from JustCloud using IE

| Location | Recovered artefacts |
|---|---|
| Network traffics | As shown in Figure 16, for downloading, JustCloud transferred the user to http://capsa.storage.googleapis.com. Connections were made with IP addresses 23.67.70.64, 74.125.133.95, 74.125.136.100, and 94.31.29.154 on port 80, 184.154.150.133 on port 443, and 184.154.150.158 on ports 80 and 443 |
| Memory | The sync name, the name of the account, and the device name |
| Registry | TypedURL and TypedURLTime respectively recorded justcloud.com's URLs and the date and the UTC time of visiting |
| Browser related files | Visited JustCloud's URLs with the date and the time of browsing in WebCacheV01.data <br> The downloading folder name and its path, the email address of the account in a URL, and the device name in json extension were seen in one of WebCacheV01.data 's tables <br> The email address of the account in %LocalAppData%\Microsoft\Internet Explorer\Recovery\Active\{4924C995-702D-11E4-9719-000C29B56A39}.dat. <br> The name of the downloaded files in: <br> • %LocalAppData%\Microsoft\Windows\INetCache\Low\IE\[Random Name]\with-others[1].htm (contained the device name as well) <br> • %LocalAppData%\Microsoft\Windows\WebCache\V01.log and WebCacheV01.dat <br> • %ProgramData%\Microsoft\Windows Defender\Support\MpWppTracing-11202014-001654-00000003-ffffffff.bin |
| MFT | "justcloud.com" word |
| NTFS Log | "justcloud.com" word and the name of the downloaded files |
| Paging file | The name of downloaded files and their saving paths |
| Unallocated space | Device name |

Table 16: Recovered artefacts associated with downloading from JustCloud using Fx.

| Location | Recovered artefacts |
|---|---|
| Network traffics | IP addresses 74.125.133.95, 94.31.29.154, 173.194.45.170 and 173.194.41.40 on port 80 and 184.154.150.133, and 184.154.150.158 on port 443. We also determined connections also took place via IP address 184.154.150.158 on port 80. |
| Memory | The email address and the password of the account (see Figure 17), the names of downloaded files and the storing path |
| Browser related files | The URLs of JustCloud with the date and the UTC time of visiting in cookies.sqlite and places.sqlite located in %AppData%\Mozilla\Firefox\Profiles\[Random Name].default\. The name of downloaded files in WebCacheV01.dat and V01.log (see Figure 18). |
| MFT and NTFS log | The name of downloaded files |
| Unallocated space | The email address of the account and the name of downloaded files with the path of storing |
| Other files | The name of downloaded files in %SystemDrive%\$Extend\$UsnJrnl:$J, %UserProfile%\ntuser.dat.LOG1 and NTUSER.DAT. justcloud.com in %ProgramFiles%\GUM5EFC.tmp\npGoogleUpdate3.dll |

Table 17: Recovered artefacts associated with downloading from JustCloud using GC.

| Location | Recovered artefacts |
|---|---|
| Network traffics | IP addresses 74.125.136.95 and 173.194.41.35 on port 80 and 184.154.150.133 and 184.154.150.158 on port 443. We also determined connections also took place via IP address 184.154.150.158 on port 80. |
| Memory | The device name and the name of downloaded files with the path of storing |
| Browser related files | History and Cookies files contained information about visited JustCloud's URLs as well as the date and the UTC time of browsing. History Provider Cache, Cookies-journal, Shortcuts, Top Sites, Top Sites-journal, Favicons, Favicons-journal, Preferences, Current Tabs, Last Session, and Current Session all located in %LocalAppData%\Google\Chrome\User Data\Default also pointed to justcloud.com. The name of the downloaded files were seen in %LocalAppData%\Google\Chrome\User Data\Default\Cache\data_1 |
| MFT, NTFS log, and paging file | They contained the names of downloaded files |
| Other files | The name of downloaded files in %ProgramData%\Microsoft\Windows Defender\Support\MpWppTracing-11202014-101334-00000003-ffffffff.bin |

Respecting the integrity check, the same result as Section 3.4 was obtained.

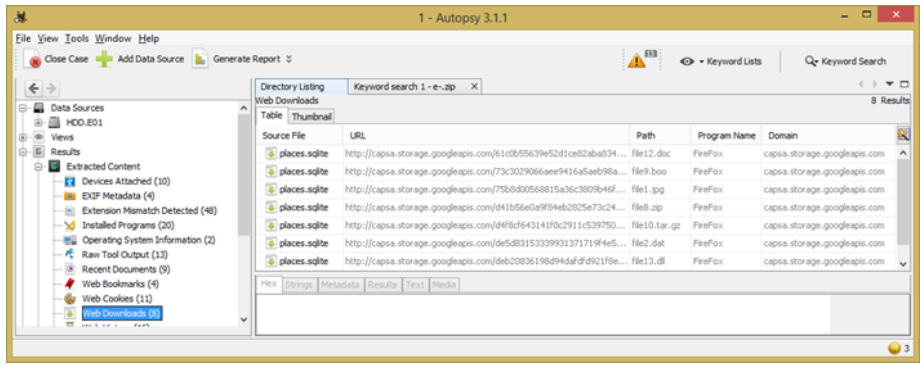

Figure 16. For downloading JustCloud used googleapis.com

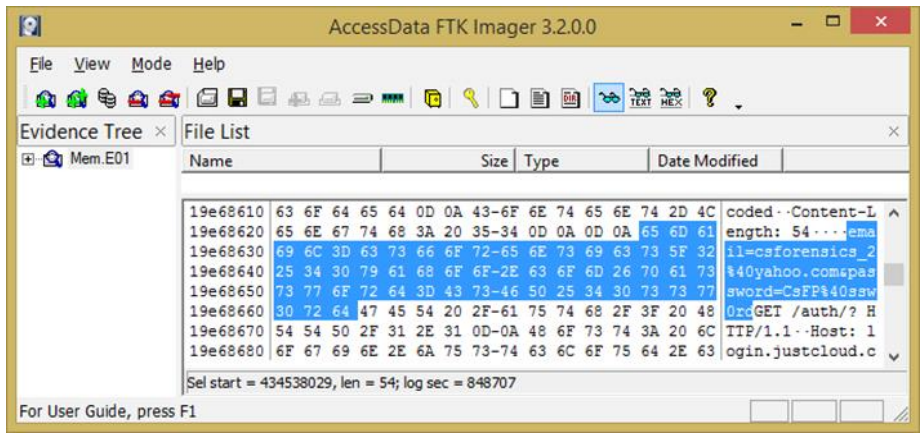

Figure 17. Email address and password in vmem file

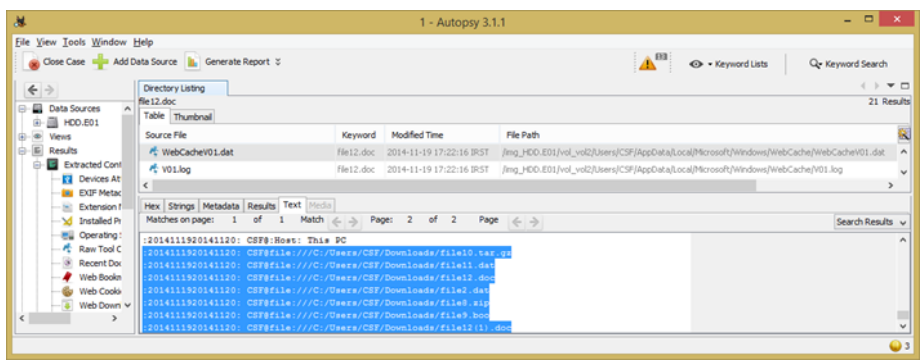

Figure 18. The name of the downloaded files and their storing paths

### 4.5 Observations: Browsing and downloading from JustCloud's iOS app

In \Apps\JustCloud\Documents, a file named syncfolder.index contained the name of the sync folder and its files. Files located in \Apps\JustCloud\Documents\datacache downloaded files with different names were observed. In \Apps\JustCloud\Library\Preferences, a file contained account's user's email was located.

## 5 Finding: pCloud

### 5.1 Observations: pCloud's account created in using IE the respective browsers

We were able to recover various information associated with the creation of the pCloud's account using the respective browsers – see Tables 18 to 20.

Table 18: Recovered artefacts associated with the creation of the pCloud's account using IE

| Location | Recovered artefacts |
|---|---|
| Network traffic | IP addresses 74.120.8.14 and 77.109.171.171 on port 80 and 74.120.8.7, 74.120.8.144, and 74.120.8.13 on port 443 |
| Memory | The email address and the password of the registered account |
| Browser related files | pCloud related addresses along with the date and the UTC time of browsing in WebCacheV01.dat<br>"pcloud" word in some files of %LocalAppData%\Microsoft\Internet Explorer\Recovery\Active\ and %LocalAppData%\Microsoft\Internet Explorer\Recovery\Last Active\ and in some subfolders of %LocalAppData%\Packages\windows_ie_ac_001\AC\INetCache\ and \Microsoft\Windows\INetCache\Low\IE\79IBA23G\qsml[1].xml |
| Registry | The URL of the visited webpages and the date and the UTC time of browsing respectively in TypedURLs and TypedURLsTime<br>The following keys were pointing to pCloud:<br>• HKEY_CLASSES_ROOT\Local Settings\Software\Microsoft\Windows\CurrentVersion\AppContainer\Storage\windows_ie_ac_001\Internet Explorer\DOMStorage\<br>• HKEY_CURRENT_USER\Software\Classes\Local Settings\Software\Microsoft\Windows\CurrentVersion\AppContainer\Storage\windows_ie_ac_001\Internet Explorer\DOMStorage\<br>• HKEY_USERS\S-1-5-21-1335463704-3291414260-3134846049-1001\Software\Classes\Local Settings\Software\Microsoft\Windows\CurrentVersion\AppContainer\Storage\windows_ie_ac_001\Internet Explorer\DOMStorage\ |
| Unallocated space | "pcloud" word |

Table 19: Recovered artefacts associated with the creation of the pCloud's account using Fx

| Location | Recovered artefacts |
|---|---|
| Network traffic | IP addresses 74.120.8.14 on port 80 and 74.120.8.14, 74.120.8.6, 74.120.8.144, and 74.120.8.13 on port 443 |
| Memory | pcloud.com and the email address and the password of the registered account |
| Browser | Files with random names indicating that pcloud.com was visited |
| related files | were located in %LocalAppData%\Mozilla\Firefox\Profiles\[Random Name].default\cache2\entries<br>Information associated with cookies and histories in cookies.sqlite and places.sqlite and the date and the UTC time of browsing identified from aforementioned files<br>"pcloud" word in permissions.sqlite, sessionstore.js, and places.sqlite-wal located in %AppData%\Mozilla\Firefox\Profiles\[Random Name].default\. The last two files also contained the email address. |
| Unallocated space | "pcloud" word and the email address of the created account |

Table 20: Recovered artefacts associated with the creation of the pCloud's account using GC.

| Location | Recovered artefacts |
|---|---|
| Network traffic | IP addresses 74.120.8.14 on port 80 and 74.120.8.14, 74.120.8.6, 74.120.8.144, and 74.120.8.12 on port 443. |
| Memory | The email address and the password of the registered account (see Figure 19) |
| Browser related files | History and Cookies files contained information about the pCloud's URLs browsed with the date and the UTC time of browsing<br>Shortcuts, Top Sites, Favicons, Current Tabs, History Provider Cache, Last Tabs and Current Session located in %LocalAppData%\Google\Chrome\User Data\Default\ pointed to pcloud.com.<br>"pcloud" word and the email address discovered in %LocalAppData%\Google\Chrome\User Data\Default\Cache\data_1 |
| MFT, NTFS log, and unallocated space | The "pcloud" word |

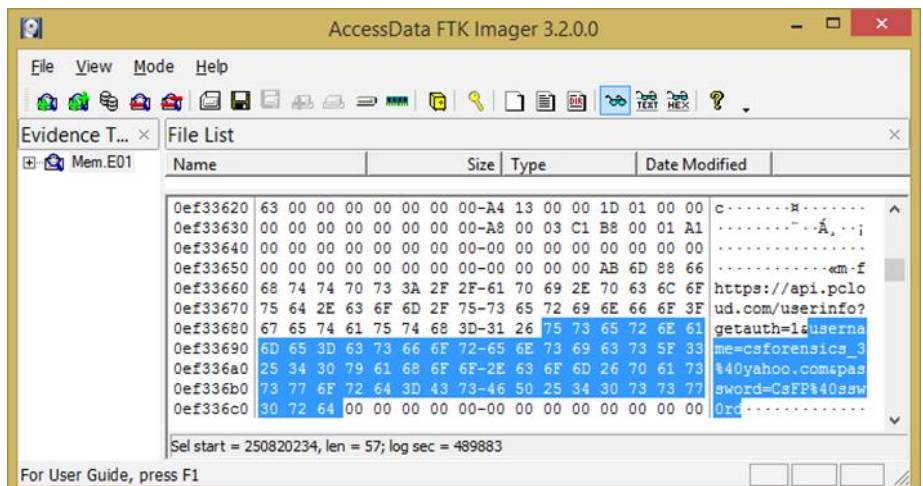

Figure 19. The email address and the password of the created pCloud account in VM-W9's vmem file

## 5.2  Observations: pCloud's Application Program

*Network traffics*. Communications between the VM and JustCloud's servers were established with IP addresses 74.120.8.17, 74.120.8.25, 74.120.8.28, 74.120.8.71, 74.120.8.74, 74.120.8.98, 74.120.8.139, 74.120.8.153, and 74.120.8.157 on port 443.
*Memory*. The email address of the account, the email address of the person who was invited to have access to the shared URL, the message of the invitation, and the share title were observed in vmem file.
*Client application files*. In %LocalAppData%\pCloud\data.db, information regarding the synced files were observed. The file was in SQLite format and contained 25 tables that described files properties and folder names of synced data. For instance, as depicted in Figure 20, the file table contained filenames and their creation and modification date and time in Unix-Numeric format. Also in the hashchecksum table, SHA1 hash values of synced files were recorded where their values matched with the original files' checksum.

Figure 20. The file table that contains filenames and their creation and modification date and time

By using Autopsy, the invited email and the invitation message were found in the same file. We also observed the invited email in data.db-wal located in the same path.
*Registry*. pCloud's default storing folder path was located in HKEY_CURRENT_USER\Software\pCloud\AppPFolders and HKEY_USERS\S-1-5-21-1335463704-3291414260-3134846049-1001\Software\pCloud\AppPFolders.
*Windows events*. In the Application node, records confirming the restore point creation and pCloud installation were located. Additionally, in Microsoft-Windows-Shell-Core%4Operational.evtx a record about pCloud was located.
*Link files*. %ProgramData%\Microsoft\Windows\Start Menu\Programs\pCloudSync\pCloud Drive.lnk pointed to pCloud's client application.
*Thumbacache files*. pCloud.exe was seen in TileCacheLogo-814531_100.dat
    The Installation filename was observed in MFT$, $LogFile, pagefile.sys, Prefetch folder, the unallocated space, and below files:

- %LocalAppData%\Temp\pCloud_Drive_20141116195829.log
- %UserProfile%\NTUSER.DAT and ntuser.dat.LOG1
- %SystemRoot%\AppCompat\Programs\Amcache.hve, Amcache.hve.LOG1, and Amcache.hve.LOG2
- %SystemDrive%\$Extend\$UsnJrnl:$J
- %SystemRoot%\System32\config\SYSTEM, SYSTEM.LOG2, SOFTWARE, and SOFTWARE.LOG2
- %LocalAppData%\IconCache.db
- %SystemDrive%\ProgramData\Package Cache\{8527342e-b5b5-4274-8f1c-01a0320d3b7d}\state.rsm

## 5.3 Observations: Uninstalling pCloud's Application Program

*Memory*. Unlike the two other CSSPs studied, pCloud automatical signing in feature is opt-in (i.e. only when a user opts to do so). Hence, no relevant evidence was located in the vmem file.

*Client application files*. Uninstall process did not remove %LocalAppData%\pCloud\data.db and, therefore, we were able to recover information mentioned in Section 5.2.

MFT, NTFS log, paging file, unallocated space, and the below files contained references to pcloud.com:

- %LocalAppData%\IconCache.db
- %LocalAppData%\Microsoft\Windows\appsFolder.itemdata-ms, appsFolder.itemdata-ms.bak, and appsFolder.itemdata-ms~RF16c076.TMP
- %LocalAppData%\Microsoft\Windows\Caches\{3DA71D5A-20CC-432F-A115-DFE92379E91F}.1.ver0x000000000000000a.db and {3DA71D5A-20CC-432F-A115-DFE92379E91F}.1.ver0x000000000000000b.db.
- %LocalAppData%\Microsoft\Windows\Explorer\TileCacheLogo-814531_100.dat
- %LocalAppData%\Temp\pCloud_Drive_20141116195829.log, pCloud_Drive_20141116195829_1_pCloud_Drive.msi.log, pCloud_Drive_20141117112055.log, and pCloud_Drive_20141117112055_0_pCloud_Drive.msi.log
- %SystemDrive%\$OrphanFiles\16a88d.rbf and license.rtf
- %SystemDrive%\$Extend\$UsnJrnl:$J
- %SystemRoot%\AppCompat\Programs\Amcache.hve, Amcache.hve.LOG1, and Amcache.hve.LOG2
- %SystemRoot%\Installer\{5950DCB0-F15C-48E0-96D9-F5948DA2A419}\pCloud.exe
- %SystemRoot%\ServiceProfiles\LocalService\AppData\Local\lastalive0.dat
- %SystemRoot%\System32\config\RegBack\SOFTWARE and SYSTEM
- %SystemRoot%\System32\config\SOFTWARE, SOFTWARE.LOG2, SYSTEM, SYSTEM.LOG1, and SYSTEM.LOG2
- %SystemRoot%\System32\wdi\{86432a0b-3c7d-4ddf-a89c-172faa90485d}\{002d5213-d0d9-41a6-ae97-4c3b05d067f2}\snapshot.etl
- %SystemRoot%\System32\wdi\{86432a0b-3c7d-4ddf-a89c-172faa90485d}\{4c28eec8-7733-4412-8c51-3088b62b54f0}\snapshot.etl

- %SystemRoot%\System32\wdi\{86432a0b-3c7d-4ddf-a89c-172faa90485d}\{f65e2d08-cfd5-4865-9d14-9b092616ddd1}\snapshot.etl
- %SystemRoot%\System32\wdi\{86432a0b-3c7d-4ddf-a89c-172faa90485d}\S-1-5-21-1335463704-3291414260-3134846049-1001_UserData.bin
- %SystemRoot%\System32\wdi\LogFiles\BootCKCL.etl
- %SystemRoot%\System32\wdi\LogFiles\StartupInfo\S-1-5-21-1335463704-3291414260-3134846049-1001_StartupInfo2.xml and S-1-5-21-1335463704-3291414260-3134846049-1001_StartupInfo3.xml
- %SystemRoot%\System32\wdi\LogFiles\WdiContextLog.etl.001, WdiContextLog.etl.002, and WdiContextLog.etl.003
- %UserProfile%\NTUSER.DAT, ntuser.dat.LOG1, and ntuser.dat.LOG2

*Registry*. The following keys pointed to pCloud:

- HKEY_CLASSES_ROOT\CLSID\{a0b73fac-351f-3948-9d8a-1dad9d870193}\InprocServer32
- HKEY_CLASSES_ROOT\CLSID\{a0b73fac-351f-3948-9d8a-1dad9d870193}\InprocServer32\1.0.0.0
- HKEY_CURRENT_USER\Software\Microsoft\Windows\CurrentVersion\UFH\SHC
- HKEY_CURRENT_USER\Software\Microsoft\Windows NT\CurrentVersion\AppCompatFlags\Compatibility Assistant\Store
- HKEY_CURRENT_USER\Software\pCloud\AppPFolders
- HKEY_LOCAL_MACHINE\SOFTWARE\Classes\CLSID\{a0b73fac-351f-3948-9d8a-1dad9d870193}\InprocServer32
- HKEY_LOCAL_MACHINE\SOFTWARE\Classes\CLSID\{a0b73fac-351f-3948-9d8a-1dad9d870193}\InprocServer32\1.0.0.0
- HKEY_LOCAL_MACHINE\SOFTWARE\Microsoft\Windows\CurrentVersion\Installer\Folders
- HKEY_USERS\S-1-5-21-1335463704-3291414260-3134846049-1001\Software\Microsoft\Windows\CurrentVersion\UFH\SHC
- HKEY_USERS\S-1-5-21-1335463704-3291414260-3134846049-1001\Software\Microsoft\Windows NT\CurrentVersion\AppCompatFlags\Compatibility Assistant\Store
- HKEY_USERS\S-1-5-21-1335463704-3291414260-3134846049-1001\Software\pCloud\AppPFolders

*Prefetch*. pcloud.exe was seen in AgAppLaunch.db, AgRobust.db, Layout.ini, MSIEXEC.EXE-B5AFA339.pf, PCLOUD.EXE-42B26121.pf, and RUNDLL32.EXE-125D4518.pf located in Prefetch folder.

*Windows events*. Records regarding the uninstallation were located in the Application node. Also pcloud.exe was observed in Microsoft-Windows-Diagnosis-Scripted%4Operational.evtx, Microsoft-Windows-Diagnostics-Performance%4Operational.evtx, Microsoft-Windows-Shell-Core%4Operational.evtx, and Microsoft-Windows-Windows Firewall With Advanced Security%4Firewall.evtx.

### 5.4 Observations: Downloading from pCloud using the respective browsers

We were able to collect information with forensic value associated with the downloading from SpiderOak using the respective browsers – see Tables 21 to 23.

Table 21: Recovered artefacts associated with downloading from pCloud using IE

| Location | Recovered artefacts |
|---|---|
| Network traffics | IP addresses 74.120.8.7, 74.120.8.12, 74.120.8.13, 74.120.8.14, 74.120.8.15, 74.120.8.86, and 74.120.8.144 on port 443 and 80.239.230.147 on 80. We also determined connections with cloudfront.net were established with IP addresses 54.230.14.77 and 54.192.15.86 on port 443. |
| Memory | The email address of the account, the invited email address, and the name of the downloaded file and its storing path |
| Registry | In addition to TypedURL, the following keys were pointing to pcloud.com:<br>• HKEY_CURRENT_USER\Software\Microsoft\Internet Explorer\LowRegistry\DOMStorage\my.pcloud.com<br>• HKEY_CURRENT_USER\Software\Microsoft\Internet Explorer\LowRegistry\DOMStorage\pcloud.com<br>• HKEY_CURRENT_USER\Software\Microsoft\Internet Explorer\LowRegistry\DOMStorage\www.pcloud.com<br>• HKEY_USERS\S-1-5-21-1335463704-3291414260-3134846049-1001\Software\Microsoft\Internet Explorer\LowRegistry\DOMStorage\my.pcloud.com<br>• HKEY_USERS\S-1-5-21-1335463704-3291414260-3134846049-1001\Software\Microsoft\Internet Explorer\LowRegistry\DOMStorage\pcloud.com<br>• HKEY_USERS\S-1-5-21-1335463704-3291414260-3134846049-1001\Software\Microsoft\Internet Explorer\LowRegistry\DOMStorage\www.pcloud.com |
| Browser related files | Visited pCloud's webpages with the date and the UTC time of browsing in WebCacheV01.data<br>The email address of the invited person in %LocalAppData%\Microsoft\Windows\INetCache\Low\IE\O0RJ133M\listshares[1].json (see Figure 21).<br>The email address of the account in the below files:<br>• %LocalAppData%\Microsoft\Windows\INetCache\Low\IE\[Random Name]\userinfo[1].json and userinfo[2].json<br>• %LocalAppData%\Microsoft\Internet Explorer\Recovery\Last Active\{3B7526CF-70B3-11E4-9719-000C29B56A39}.dat<br>Browsed uploaded filenames in the following file with the date and the time of creation and modification:<br>• %LocalAppData%\Microsoft\Windows\INetCache\Low\IE\[Random Name]\listfolder[1].json<br>Several js and json files located in the above path as well as the below files were pointing to pcloud.com:<br>• %LocalAppData%\Microsoft\Windows\INetCookies\Low\ZY2FBEQ8.txt<br>• %LocalAppData%\Microsoft\Windows\WebCache\V01.log<br>• %UserProfile%\NTUSER.DAT and ntuser.dat.LOG1 |
| MFT and the unallocated space | The "pcloud.com" and the email address of the account with the date of registration (see Figure 22) |

Table 22: Recovered artefacts associated with downloading from pCloud using Fx

| Location | Recovered artefacts |
|---|---|
| Network traffics | IP addresses 74.120.8.6, 74.120.8.7, 74.120.8.12, 74.120.8.15, 74.120.8.86, and 74.120.8.144 on 443, and IP address 74.120.8.15 on port 80. We also determined that communications with cloudfront.net were established on IP addresses 54.230.26.16, 54.230.26.91, 54.230.229.182, and 54.230.229.204 on port 443. |
| Memory | The name of the downloaded file along with its storing path, the email address of the account, account creation date and time, the shared folder, its recipient and access permissions, and the date and the time of sharing (see Figure 23) |
| Browser related files | Visited pCloud's URLs with the date and the UTC time of browsing in cookies.sqlite and places.sqlite located in %AppData%\Mozilla\Firefox\Profiles\[Random Name].default\ <br> The name of the downloaded file in places.sqlite and its moz_annos table <br> The email address of the account, the shared folder name, access permissions, and the recipient in some files located in %LocalAppData%\Mozilla\Firefox\Profiles\[Random Name].default\cache2\entries\[RandomFolderName]\ and %LocalAppData%Mozilla\Firefox\Profiles\[Random Name].default\cache2\trash14488\ |
| MFT, NTFS log, and pagefile.sys | pcloud.com and the name of the downloaded files and its saving path. |
| Link files | %AppData%\Microsoft\Windows\Recent\archivedwl-740.lnk pointing to a downloaded archive that contained all sample files |
| Unallocated space | The email address of the account with the date of account registration, the names of downloaded files, the shared folder names, access permissions, and their recipients |
| Other files | The names of downloaded files in: <br> • %AppData%\Microsoft\Windows\Recent\CustomDestinations\6824f4a902c78fbd.customDestinations-ms. <br> • %LocalAppData%\Microsoft\Windows\WebCache\ WebCacheV01.dat and V0100036.log <br> • %SystemDrive%\$Extend\$UsnJrnl:$J <br> • %UserProfile%\ NTUSER.DAT, ntuser.dat.LOG1, and ntuser.dat.LOG2 <br> • %ProgramData%\Microsoft\Search\Data\Applications\Windows\GatherLogs\SystemIndex\SystemIndex.3.gthr <br> Also pcloud.com was seen in two mum files located in %SystemDrive%\$OrphanFiles\. |

Table 23: Recovered artefacts associated with downloading from pCloud using GC

| Location | Recovered artefacts |
|---|---|
| Network traffics | IP addresses 74.120.8.6, 74.120.8.7, 74.120.8.13, 74.120.8.14, 74.120.8.15, 74.120.8.86, and 74.120.8.144 on 443 were identified as pCloud servers' IPs. Connections to 74.120.8.14 were made on port 80, and communications established with cloudfront.net took place via IP addresses 54.230.26.21 and 54.230.26.96 on port 443. |
| Memory | The name of the downloaded file and its storing path, the email address of the account, the password of the account, the creation date and time of the account, the name of the shared folder, the recipient of the shared folder, permissions, and the date and the time of sharing |
| Browser related files | History and Cookies files contained information about visited pCloud's webpages with the date and the UTC time of browsing. The name of the downloaded file in History. History Provider Cache, Cookies-journal, Shortcuts, History-journal, Top Sites, Top Sites-journal, Favicons, Favicons-journal, Preferences, Current Tabs, and Current Session, all located in %LocalAppData%\Google\Chrome\User Data\Default were pointing to pcloud.com. The email address of the account and the invited email addresses in %LocalAppData\Local\Google\Chrome\User Data\Default\Cache\data_1 |
| MFT and NTFS log | The name of the downloaded file. |
| Other files | The name of the downloaded file in $Extend\$UsnJrnl:$J. |

We determined that the MD5 checksum of sample files, except file13.dll, were not altered after downloading. However, we were not able to restore ADS from file13.dll. Metadata of file12.doc such as Authors value remained the same as the original file. It was also observed that Modified value of downloaded files was the date of uploading. The values of Created and Accessed fields were, on the other hand, changed to the date and the time of downloading.

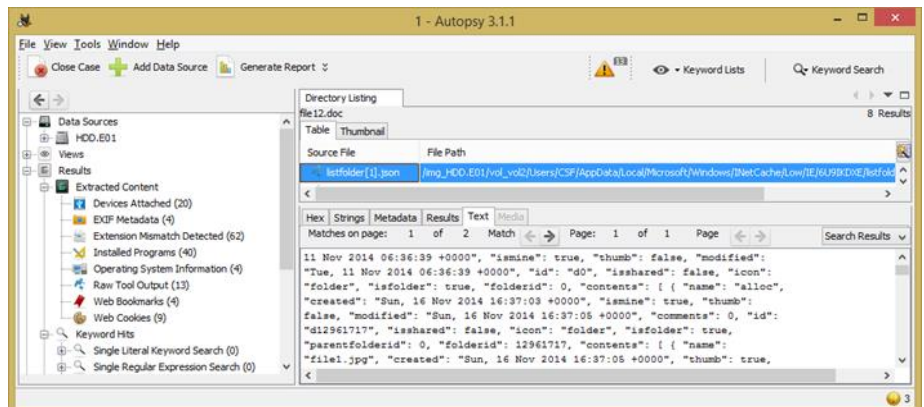

Figure 21. The invited email address in a json file

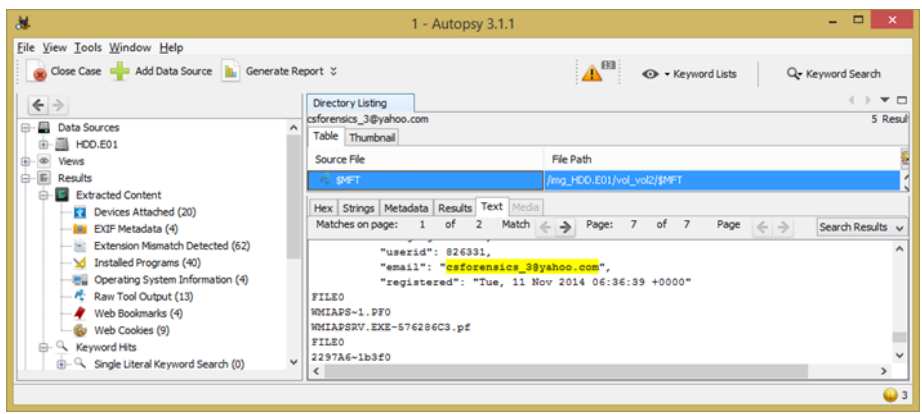

Figure 22. Account email with the date of registration

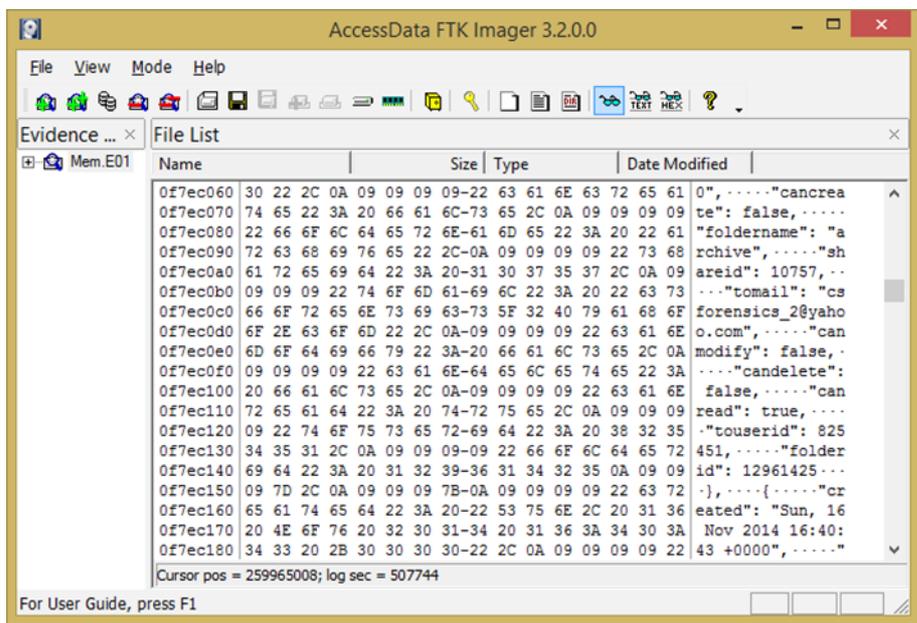

Figure 23. The email address of the account with the date of registration

### 5.5  Observations: Browsing and downloading from pCloud's iOS app

A file named p.db located in \Apps\pCloud\Library\Application Support contained table and data similar to those discussed in Section 5.2. We located file12.doc as one of the files opened in the pCloud app in \App\pCloud\Library\Caches\com.pcloud.pcloud.cache. Although the name of the located file was different, the MD5 signature indicated that the contents of the file had not been changed. The metadata of the file was identical to the original one. The only other sample file found in the pCloud's app folder was file1.jpg, but had a different name and MD5 hash. Due to the lack of collected data, we were not able to draw a conclusion regarding the integrity verification in pCloud's mobile application.

## 6  Conclusion and Future Work

In this research, we located and described various artefacts of forensic when SpiderOak, JustCloud, and pCloud were used with IE, Fx, and GC browsers, client application, and mobile application on Windows machines and iOS devices. The recovered artefacts include email addresses, the ID, and the name of the created account and the name of the uploaded and downloaded files. Our findings also suggested that user's credentials could be recovered from memory, and the checksums of sample files after being downloading from investigated CSSPs remained identical to the original. However, we noted that none of the investigated CSSPs could prevent the timestamp and the alternate data streams (ADS) of files from being changed. Our findings also indicated that metadata of the doc file examined in this study was not altered which could be another piece of useful information in forensic investigation.

Future work would extending our work to examining machines running Linux and other less popular operation systems.